\documentclass[footinbib, aps,pra,reprint, 10pt,twocolumn,showpacs, superscriptaddress]{revtex4-1}
\usepackage{amsfonts,amsmath,amssymb}%, siunitx}
\usepackage{blindtext}
\usepackage{graphicx}
\usepackage[pdfstartview=FitH,colorlinks=true,linkcolor=blue,citecolor=blue,urlcolor=blue]{hyperref}
\usepackage{tikz}
\usepackage{verbatim}
\usepackage{comment}
\usepackage{times}

\usepackage[english]{babel}
\usepackage{latexsym}
\usepackage{graphics}
\usepackage{subfigure}
\usepackage{epsfig}
\usepackage{color}
\usepackage{braket}
\usepackage[T1]{fontenc}
\usepackage[latin9]{inputenc}
\usepackage{amstext}
\usepackage{esint}
\usepackage[normalem]{ulem}   % needed for striking out text with \sout

\newcommand{\beq}{\begin{equation}}
\newcommand{\eeq}{\end{equation}}

\begin{document}

\title{Correlation properties of a one-dimensional repulsive Bose gas at finite temperature}

\author{Giulia De Rosi}
\email{giulia.de.rosi@upc.edu}
\affiliation{Departament de F\'isica, Universitat Polit\`ecnica de Catalunya, Campus Nord B4-B5, 08034 Barcelona, Spain}

\author{Riccardo Rota}
%\email{riccardo.rota@epfl.ch}
\affiliation{Institute of Physics, Ecole Polytechnique F\'ed\'erale de Lausanne (EPFL), CH-1015 Lausanne, Switzerland}

\author{Grigori E. Astrakharchik}
\email{grigori.astrakharchik@upc.edu}
\affiliation{Departament de F\'isica, Universitat Polit\`ecnica de Catalunya, Campus Nord B4-B5, 08034 Barcelona, Spain}
\affiliation{Departament de F{\'i}sica Qu{\`a}ntica i Astrof{\'i}sica, Facultat de F{\'i}sica, Universitat de Barcelona, E-08028 Barcelona, Spain}
%{\bf 3} Institut de Ci{\`e}ncies del Cosmos, Universitat de Barcelona, ICCUB, Mart{\'i} i Franqu{\`e}s 1, E-08028 Barcelona, Spain\\

\author{Jordi Boronat}
\email{jordi.boronat@upc.edu}
\affiliation{Departament de F\'isica, Universitat Polit\`ecnica de Catalunya, Campus Nord B4-B5, 08034 Barcelona, Spain}

\date{\today}

\begin{abstract}
We present a comprehensive study shedding light on how thermal fluctuations affect correlations in a Bose gas with contact repulsive interactions in one spatial dimension. The pair correlation function, the static structure factor, and the one-body density matrix are calculated as a function of the interaction strength and temperature with the exact ab-initio Path Integral Monte Carlo method. We explore all possible gas regimes from weak to strong interactions and from low to high temperatures. 
We provide a detailed comparison with a number of theories, such as perturbative (Bogoliubov and decoherent classical), effective (Luttinger Liquid) and exact (ground-state and thermal Bethe Ansatz) ones.
Our Monte Carlo results exhibit an excellent agreement with the tractable limits and provide a fundamental benchmark for future observations which can be achieved in atomic gases, cavity quantum-electrodynamic and  superconducting-circuit platforms. 
\end{abstract}

\maketitle

\section{Introduction}

Understanding the role of strong correlations in complex quantum many-body systems such as gases of interacting atoms or electrons is one of the most important challenges in modern condensed matter physics, material research, and chemistry. In this context, one-dimensional (1D) Bose gases composed by particles with contact repulsive interactions \cite{Mistakidis2022} have proven to be a versatile testbed for the study of strong correlations in quantum many-body physics, where solid-state, low-temperature, and atomic physics converge. They offer a simple yet nontrivial model which can still be captured with reasonable theoretical effort \cite{Cazalilla2011}, by allowing for the direct comparison between exact and analytical results.

The 1D Bose gas model may be precisely simulated by cavity quantum-electrodynamic devices, by shedding light on the behavior of its correlation properties \cite{Barrett2013}. Most importantly, it has been experimentally realized in superconducting circuits where correlations have been indeed measured from weakly to strongly interacting regime \cite{Eichler2015}.
However, the best experimental platform so far was provided by ultracold atoms as they offered an exquisite and precise control over many system parameters \cite{Bloch2008}, including the interaction strength between atoms. These ultracold atomic gases are so dilute that the in-depth understanding of the physics at stake is possible in most cases. Finally, they can be realized in different spatial dimensions, including 1D geometry.

At finite temperature, many momentum modes in the longitudinal 1D direction are excited. This leads to markedly different behavior of a 1D Bose gas compared to the three-dimensional Bose-Einstein condensate (BEC), where the lowest momentum mode remains macroscopically occupied at low temperature. The excitation of momentum modes with temperature is the origin of highly enhanced density and phase fluctuations which prevent the off-diagonal long-range order \cite{Mermin1966, Hohenberg1967} and then the formation of a true BEC. Instead, an extraordinarily rich landscape of many different degenerate regimes emerges \cite{Kheruntsyan2003, Sykes2008, Deuar2009, Vogler2013, Salces-Carcoba2018, DeRosi2019, DeRosi2021II}, which are separated by smooth crossovers and they might or not share the typical features of a BEC. They can be explored by changing the interaction strength and temperature.

Distinct quantum regimes can be conveniently classified by the characteristic behavior of thermodynamic properties. Asymptotic analytical limits have been described by a variety of physical models \cite{DeRosi2019, DeRosi2021II} and compared with the exact solution provided by the thermal Bethe-Ansatz (TBA) method within the Yang-Yang theory \cite{Yang1969, Yang1970}. A new intriguing quantum regime in a 1D Bose gas has been recently predicted by the same authors of the present manuscript \cite{DeRosi2021II}. It is signalled by a thermal anomaly in the temperature dependence of the specific heat for any finite value of the interaction strength. Calculations of the dynamic structure factor with the ab-initio Path Integral Monte Carlo (PIMC) method showed that, at temperatures similar to the anomaly threshold, the excitation pattern experiences the breakdown of the quasiparticle description holding instead at low temperature \cite{DeRosi2021II}. Both behaviors of the specific heat and the dynamic structure factor suggest that the novel so-called hole anomaly, which is induced by the presence of unpopulated states in the excitation spectrum of the system, is a reminiscence of the superfluid-normal phase transition at the critical temperature, although any phase transition is not allowed in 1D geometry \cite{Landau2013}.

Different quantum regimes in a 1D Bose gas can be also identified with the corresponding analytical limits of the pair correlation function (also called two-body distribution function) \cite{Kheruntsyan2003, Sykes2008, Deuar2009}. However, the full function cannot be readily obtained with TBA and require instead ab-initio techniques.
At zero temperature, several correlation properties have been calculated for arbitrary interaction strength with quantum Monte Carlo \cite{Astrakharchik2003, Astrakharchik2006} and density matrix renormalization group (DMRG) \cite{Verstraete2010} techniques.
Previous finite-temperature studies only focus on very limited regimes of interaction and temperature and are based on different approaches. In the weakly-interacting limit, some mean-field methods, including Bogoliubov theory \cite{Mora2003, Castin2004}, and the stochastic Gross-Pitaevskii equation \cite{Henkel2017} can be employed at low temperature, while the semiclassical expansion \cite{Bastianello2020} is valid only at high temperature.
In the strongly-interacting regime, available methods are the Bose-Fermi mapping~\cite{Cherny2006} and the numerical stochastic gauge technique whose results are only reported for a single value of temperature~\cite{Drummond2004}.
At temperature below the hole anomaly, correlation properties have been computed with Bethe-Ansatz method~\cite{Panfil2014}, PIMC technique using worm updates~\cite{Xu2015}, by relying on the mapping with the sinh-Gordon model~\cite{Kormos2009, Kormos2010} and on the form factor approach \cite{Cheng2022}. 
At high temperature, imaginary time stochastic gauge-$P$ simulations can be employed~\cite{Deuar2009}. Low-momentum~\cite{DeNardis2016} and large-distance~\cite{Cazalilla2004, Kozlowski2011, Patu2013, Klumper2014} correlations have been also explored.
% Brandt method?

The present work fills an important existing gap by providing a complete study of the stationary microscopic correlation properties of a 1D Bose gas for a wide range of interaction strength, all over from the weakly- to the strongly-repulsive regime, and for temperatures crossing from the quantum to the classical gas.
We calculate the spatial dependence of the one-body density matrix and the pair correlation function, as well as the static structure factor, as a function of the momentum in a very broad range of values. This calculation is carried out using the PIMC technique which allows us to obtain unbiased results with a controllable accuracy for any temperature and interaction strength values. Our results show an excellent agreement with exact thermal Bethe-Ansatz solution \cite{Yang1969, Yang1970} and analytical limits in regimes of their validity.
Our PIMC findings are fundamental for benchmarking the theoretical predictions and to stimulate future experimental measurements.

The structure of the paper is as follows. In Sec.~\ref{Sec:Model} we introduce the studied model of the 1D Bose gas at finite temperature. Section~\ref{Sec:PIMC} is devoted to a brief description of the Path Integral Monte Carlo method used for the numerical results of the correlation functions. Our findings of the pair correlation function, the static structure factor, and the one-body density matrix are reported in Secs.~\ref{Sec:g2}, \ref{Sec:S(k)}, and \ref{Sec:OBDM}, respectively, and they exhibit an excellent agreement with analytical and thermal Bethe-Ansatz limits. In Sec.~\ref{Sec:experiments}, we discuss possible upcoming experimental observations of our predictions. Finally, in Sec.~\ref{Sec:conclusions}, we draw the conclusions and future perspectives of our work.

\section{Model}
\label{Sec:Model}

The many-body Hamiltonian of a 1D homogeneous gas of $N$ Bose particles interacting via repulsive contact pseudopotential (Lieb-Liniger model) is given by
\begin{equation}
\label{Eq:H}
H = - \frac{\hbar^2}{2m}\sum_{i = 1}^N \frac{\partial^2}{\partial x_i^2} + g\sum_{i > j}^N\delta(x_i - x_j),
\end{equation}
where $m$ is the atomic mass, $g = - 2 \hbar^2/(m a)>0$ is the positive 1D coupling constant~\cite{Olshanii1998}, and $a<0$ is the negative 1D $s$-wave scattering length. The dimensionless interaction strength $\gamma=-2/(n a)$ is related to the gas parameter $n a$, where $n=N/L$ is the linear density with $L$ being the length of the system. We are interested here in the thermodynamic properties which are obtained by taking the $N, L \to  \infty$ limit while keeping the density $n$ constant. There is a continuous crossover which encompasses different quantum degeneracy regimes. In the Gross-Pitaevskii (GP) limit of weak repulsion $\gamma \ll 1$ and high density $n |a| \gg 1$ the gas admits a mean-field description \cite{Pitaevskii2016}. In the Tonks-Girardeau (TG) regime \cite{Tonks1936, Girardeau1960} of very strong repulsion $\gamma \gg 1$ and low density $n |a| \ll 1$, bosons become impenetrable and they cannot cross each other. This constraint, together with the spatial peculiarity of 1D systems, acts as an effective Pauli exclusion principle, resulting in a dramatic suppression of three-body losses and making the TG gas stable. In this highly correlated limit, the wavefunction of strongly repulsive bosons can be mapped onto that of an ideal Fermi gas, resulting in identical thermodynamics and excitation spectrum \cite{Girardeau1960}. Seminal experiments have explored the GP-TG crossover in the past years \cite{Paredes2004, Kinoshita2004, Tolra2004, Kinoshita2005, Haller2011, Jacqmin2011, Guarrera2012}.

At zero temperature, the Bethe-Ansatz method allows one to calculate exact energetic properties of the Lieb-Liniger model such as the ground-state energy $E_0$, chemical potential $\mu_0 = (\partial E_0/\partial N)_{a, L}$ and sound velocity $v = \sqrt{n/m (\partial \mu_0/\partial n)_{a}}$ which are all functions of the interaction strength $\gamma$ \cite{Lieb1963, Lieb1963II, Pitaevskii2016, DeRosi2017}.
The speed of sound $v$ monotonically increases with $\gamma$ from the mean-field $v_{\rm GP} =\sqrt{g n/m}$ to the Fermi $v_F = \hbar \pi n/m$ value in the TG limit.

At finite temperature, the complete thermodynamics within the canonical ensemble can be inferred from the Helmholtz free energy $A = E - TS$, where $E$ is the internal energy and $S = - \left(\partial A / \partial T \right)_{a, N, L}$ the entropy.
Its knowledge can be used to calculate the pressure
\begin{equation}
\label{Eq:Pressure}
P=-(\partial A/\partial L)_{T,a,N} ,
\end{equation}
and the inverse isothermal compressibility
\begin{equation}
\label{Eq:kTINV}
\kappa_T^{-1} = \left( \partial P/\partial n  \right)_{T, a, N} = n \left( \partial \mu/\partial n  \right)_{T, a, N} 
\end{equation}
where $\mu = \left( \partial A/\partial N  \right)_{T, a, L}$ is the chemical potential at finite temperature and we have employed the Gibbs-Duhem relation $dP = n d\mu + s dT$ with $s = S/L$ the entropy density. 
The quantity $\left( \partial n/\partial \mu  \right)$ is directly related to the atom number fluctuations which can be measured in ultracold atom experiments. 
In this way, one has access to the isothermal compressibility, $\kappa_T$, equation of state and temperature $T$ \cite{Gemelke2009, Sanner2010, Muller2010, Jacquim2011, Hartke2020}. 
The experimental estimate of temperature is also possible through the use of Eq.~\eqref{Eq:kTINV} and the combination of the measurements of the density $n$, the isothermal compressibility $\kappa_T$ and the pressure $P$ \cite{Ku2012, Desbuquois2014}.

In a system with zero-range interaction, one can define the Tan's contact parameter~\cite{Olshanii2003,Braaten2011,Yao2018, DeRosi2019}
\begin{equation}
\label{Eq:contact}
\mathcal{C}= \frac{4 m}{\hbar^2} \left(\frac{\partial A}{\partial a}  \right)_{T,N,L}  ,
\end{equation}
which connects short-range (large-momentum) correlations with the thermodynamic functions \cite{Olshanii2003,Tan2008,Tan2008II,Tan2008III, Barth2011}. A set of exact thermodynamic relations holding for any value of interaction strength and temperature, and based on simple scaling considerations \cite{FetterBook, Barth2011}, has been recently derived \cite{DeRosi2019}.

\section{Path Integral Monte Carlo Technique}
\label{Sec:PIMC}

The PIMC is a powerful stochastic method which allows to obtain numerically structural and energetic properties of a quantum system at finite temperature \cite{Ceperley1995} which is described here by the microscopic Hamiltonian~\eqref{Eq:H}. Some information on the structure of the excitation spectrum can be inferred as well from the dynamic structure factor by using the inverse Laplace transform \cite{DeRosi2021II}. PIMC method relies on the Feynman path-integral description of an ensemble of $N$ quantum atoms in terms of a set of $N$ classical polymers, each of them reproducing a quantum delocalized particle~\cite{Ceperley1995}. In this way, the thermal average or expectation value $\langle O \rangle$ of the quantum observable $O$ is expressed as a multidimensional integral which can be efficiently computed via Monte-Carlo sampling of the coordinates $R$:
\begin{equation}
\label{Eq:average}
\langle O \rangle =  \textrm{Tr} \left(n_T O \right) = \frac{1}{Z N!} \sum_\mathcal{P} \int dR \ G(R,\mathcal{P} R;\beta) O(R) 
\end{equation}
where we have introduced the thermal density matrix $n_T = e^{-\beta H}/Z$, the partition function $Z = \textrm{Tr} \left(e^{-\beta H}\right)$, the inverse temperature $\beta = \left(k_B T  \right)^{-1}$ and $H$ is the Hamiltonian, Eq.~\eqref{Eq:H}.
In the following, we work in coordinate representation $G(R_1,R_2;\beta) = \langle R_2 | e^{-\beta H} | R_1\rangle$ and $O(R)= \langle R|O|R\rangle$ where $R_i = \{x_{1,i}, x_{2,i} \ldots, x_{N,i} \}$ denotes a set of the coordinates of the $N$ atoms of the system and $G(R_1,R_2;\beta)$ is the Green function propagator describing the evolution in the imaginary time $\beta$ from the initial $R_1$ to the final $R_2$ configuration.
The configuration $\mathcal{P} R$ in Eq.~\eqref{Eq:average} is obtained by applying a permutation $\mathcal{P}$ of the particle labels to the initial configuration $R$ and the sum $\sum_\mathcal{P}$ over the $N!$ permutations allows to take into account the quantum statistics of the identical bosonic atoms.

The key aspect of the Path Integral formalism is the convolution property of the propagator
\begin{equation}
\label{Eq:convolution}
G(R_1,R_3;\beta_1+\beta_2) = \int \, dR_2 G(R_1,R_2;\beta_1) G(R_2,R_3;\beta_2)\;,
\end{equation}
which can be generalized to a series of intermediate steps $R_2 \ldots R_M$ (technique known as {\em trotterization}) defining a path with $M$ configurations and total time $\beta = \varepsilon M$, where $\varepsilon$ is the time step.
For a finite value of $M$, the path is discrete in time. In the limit of large $M$, the path becomes continuous and
$\varepsilon$ approaches zero, so each step corresponds to high temperatures $T$. In this classical gas limit, the propagator admits an analytical approximation where the quantum effects of the non-commutativity between the kinetic and interaction potential operators in the Hamiltonian are neglected. The thermal expectation value, Eq.~\eqref{Eq:average}, can be then approximated as
\begin{equation}
\label{Eq:QuantumAverage}
\langle O \rangle \simeq \frac{1}{Z N!} \sum_\mathcal{P} \int  dR_1 \ldots dR_M O(R_1) \prod_{i = 1}^M G(R_i,R_{i+1};\varepsilon)
\end{equation}
where the Bose-Einstein statistics imposes the boundary condition $R_{M+1} = \mathcal{P} R_1$. The probability distribution $p(R_1,\ldots,R_M) =  \prod_{i=1}^M G(R_i,R_{i+1};\varepsilon)$ is positive definite and its integral over the space of configurations is equal to the unity. The PIMC approach is based on a stochastic evaluation of the integral in Eq.~\eqref{Eq:QuantumAverage} by sampling $N \times M$ degrees of freedom according to the probability distribution $p(R_1,\ldots,R_M)$. In order to improve the permutation sampling, we use the worm algorithm~\cite{Boninsegni2006}. Decomposition~\eqref{Eq:QuantumAverage} becomes exact in the limit $M \to \infty$ where the imaginary time $\varepsilon$ is small and the analytical high-temperature approximation for the propagator $G$ is accurate, allowing for an exact calculation of the thermal averages $\langle O \rangle$ with the PIMC method. We approximate the propagator within a pair-product scheme which is based on the exact solution of the two-body problem for the Hamiltonian in Eq.~\eqref{Eq:H} \cite{Gaveau1986,Yan2015}. The optimization of the number of the convolution terms $M$ in Eq.~\eqref{Eq:QuantumAverage} has been achieved by benchmarking the PIMC expectation values of the internal energy per particle $E/N$ and the isothermal compressibility $\kappa_T$ as a function of temperature and interaction strength against the exact thermal Bethe-Ansatz results \cite{DeRosi2021II}.

The PIMC method requires an average time of several days for performing a single calculation as the computational cost scales quadratically with the number of particles $N$. This scaling is much more favorable as compared to the diagonalization methods which typically have an exponential cost. 
PIMC thus allows us to efficiently calculate various observables, including correlation functions, by obtaining a mean value and an error bar. The errors are characterized by two different contributions. The first one is the statistical error which can be reduced by increasing the number of iterations. It typically scales with the inverse of the square root of the number of iterations. The second one is the systematic error which is controlled by decreasing the time step $\varepsilon$ and increasing the number of configurations $M$. 
By fixing $\varepsilon$, it is easy to see that $M$ is proportional to the inverse of temperature $M \sim \beta \sim 1/T$. 
The lower is the temperature, the larger is the required $M$ to keep the systematic error sufficiently small. 
In the classical limit of high temperatures, the effects of quantum delocalization of particles are less relevant and a small value of $M$ is sufficient for an accurate description of the many-body system.
In the quantum regime of low temperatures, it is necessary instead a large $M$ which makes the calculation more time demanding. 
By fixing the temperature, the computational cost also depends on the interaction strength $\gamma$ which scales with the inverse of the linear density $\gamma \sim 1/n$ where $n \sim N$. In the Gross-Pitaevskii regime of weak interactions $\gamma \ll 1$ and high density, a large number of atoms $N$ is needed, making calculations challenging. By approaching the opposite Tonks-Girardeau regime of strong interactions $\gamma \to \infty$, $N$ can be relatively small. 
However, the importance of quantum correlations is enhanced by increasing $\gamma$, thus requiring a large value of $M$. 
To summarize, within the PIMC calculations, it is possible to keep the error bars under control and to decrease them in a systematic way at the expenses of longer simulation times.

In the following Sections, we report the PIMC results for the pair correlation function, the static structure factor, and the one-body density matrix, whose expectation values have been computed with  Eq.~\eqref{Eq:QuantumAverage} for a broad range of values of interaction strength $\gamma$ and temperature.

\section{Pair correlation function}
\label{Sec:g2}
The pair correlation function or normalized density-density correlator quantifies the probability of finding two particles separated by a distance $x$ \cite{Pitaevskii2016}, according to 
\begin{equation}
\label{Eq:g2}
g_2 \left(x = x_1 - x_2 \right) = \frac{\langle  \hat{\psi}^\dagger\left(x_2\right)   \hat{\psi}^\dagger\left(x_1\right) \hat{\psi}\left(x_1\right)\hat{\psi}\left(x_2\right) \rangle}{n^2}
\end{equation}
where $\hat{\psi}\left(x \right)$ is the bosonic field operator and $\langle\cdots\rangle$ denotes an average over an ensemble at thermal equilibrium at a given temperature $T$.
Here $n = \langle  \hat{\psi}^\dagger \hat{\psi}\rangle$ is the diagonal density where the operators are evaluated at the same spatial coordinate. The pair correlation function provides the characteristic length scale over which the density-density fluctuations decay.
In a 1D Bose gas, $g_2\left(x\right)$ plays a key role in the study of interference properties of atom lasers in a 1D waveguide \cite{Guerin2006}, by fostering practical applications such as matter-wave interferometry \cite{Hofferberth2007}.
In the experimental image of an expanding gas cloud, $g_2\left(x\right)$ can be employed to probe complex many-body states of trapped ultracold atoms \cite{Altman2004}.

The local pair correlation function $g_2\left(0\right)$, at $x = 0$, gives the probability of two particles to overlap. Its value is related to the derivative of the free energy with respect to the coupling constant $\left(\partial A/\partial g\right)_{T, N, L}$ \cite{Kheruntsyan2003} according to the Hellmann-Feynman theorem \cite{Feynman1939}. By using Eq.~\eqref{Eq:contact} and the definition of the interaction strength $\gamma$, one finds the relation with the Tan's contact $\mathcal{C}$ \cite{Pitaevskii2016}:
\begin{equation}
\label{Eq:g20}
g_2 \left( 0  \right) = \frac{1}{\gamma^2 n^3 } \frac{\mathcal{C}}{N}\;.
\end{equation}
Its value can be calculated exactly in a 1D Bose gas through the thermal Bethe Ansatz \cite{Yang1969, Yang1970, Kheruntsyan2003, DeRosi2019}. Since $g_2\left(0\right)$ and $\mathcal{C}$ are proportional, Eq.~\eqref{Eq:g20},  they both connect the short-range behavior of correlations with the thermodynamic properties. 
Analytical limits of $g_2\left(0 \right)$ and $\mathcal{C}$ in various physical regimes are reported in Refs.~\cite{Kheruntsyan2003, Gangardt2003II} and \cite{DeRosi2019}, respectively.

At large relative distances in a gas phase, atoms become uncorrelated and the pair correlation function approaches the $g_2 \to 1$ value.

In the decoherent classical (DC) regime, an excellent approximation for the pair correlation function is \cite{Sykes2008, Deuar2009}
\begin{equation}
\label{Eq:g2 DC}
g_2\left(x\right)_{\rm DC} = 1 + e^{- x^2/\sigma^2} - \gamma n \sigma  \sqrt{\pi} \text{erfc}\left( \frac{|x|}{\sigma} \right)
\end{equation}
where $\sigma = \lambda/\sqrt{2 \pi}$ is proportional to the thermal de Broglie wavelength $\lambda = \sqrt{2 \pi \hbar^2/\left( m k_B T \right)}$ and $\text{erfc}\left(x\right)$ is the complementary error function. The DC regime holds for $k_B T\gg\text{max}\left\{2\hbar^2/(ma^2), \hbar^2n^2/(2m) \right\}$.
That is, the temperature must be large compared both to the characteristic energy associated with the $s$-wave scattering and to the quantum degeneracy temperature
$T_d = T_F/\pi^2$ related to the Fermi energy $E_F = k_B T_F = \hbar^2 \pi^2 n^2/\left(2 m\right)$. 
For $T \gg T_d$, the chemical potential $\mu$ is large and negative, so the bosonic occupation number is small $n(k) \ll 1$ and can be approximated by the Maxwell-Boltzmann distribution $n(k)_{\rm MB}  \approx e^{-\beta\left[\hbar^2 k^2/(2m) -\mu \right]}$, where $k$ is the momentum. Such an approximation yields Eq.~\eqref{Eq:g2 DC}. In the DC regime, both phase and density fluctuations are large, thermal effects always dominate over the interactions and the system approaches the ideal Bose gas behavior at very high temperature even if quantum effects still play a role \cite{Kheruntsyan2003, Sykes2008, Deuar2009, DeRosi2021II}, as signalled by the corrective term in Eq.~\eqref{Eq:g2 DC} depending on the interaction strength $\gamma$. The parameter $\sigma$ is thus always smaller than the absolute value of the 1D $s-$wave scattering length: $\sigma < |a|$.

In the non-interacting limit $\gamma \to 0$ of Eq.~\eqref{Eq:g2 DC}, the classical ideal gas result~\cite{Naraschewski1999} is recovered and characterized by the Gaussian Maxwell-Boltzmann decay at large distance with a correlation length fixed by $\sigma$ which determines the approaching to the uncorrelated limit $g_2 \to 1 $.

\begin{figure}[h!]
\includegraphics[width=0.5\textwidth]{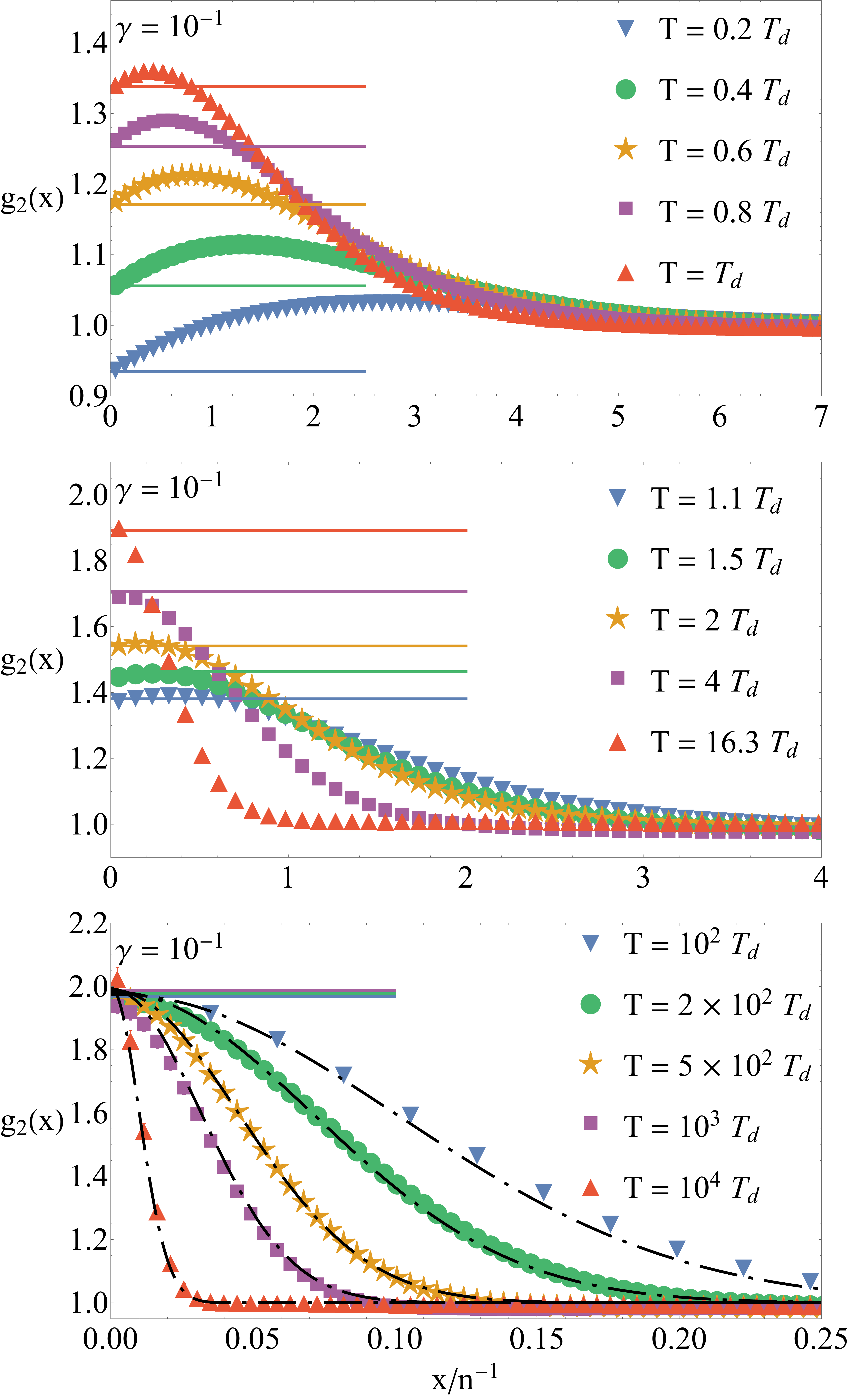}
\caption{Pair correlation function $g_2\left(x\right)$ as a function of the interatomic relative distance $x$ in units of the inverse density $n^{-1}$ for the interaction strength $\gamma = 10^{-1}$. Different values of temperature in units of the quantum degeneracy value $T_d = \hbar^2 n^2/\left(2 m k_B\right)$ are reported. Symbols denote exact Path Integral Monte Carlo results and their sizes are larger than the statistical error bars. Horizontal lines represent the local pair correlation function $g_2\left(0\right)$ calculated with thermal Bethe Ansatz, Eq.~\eqref{Eq:g20}, and they are reported in an increasing order of the temperature from low (bottom) to high (top) values in each panel. In the lowest panel, black dot-dashed lines show the decoherent classical limit, Eq.~\eqref{Eq:g2 DC}.
}
\label{fig:g201}
\end{figure}

\begin{figure}[h!]
\includegraphics[width=0.5\textwidth]{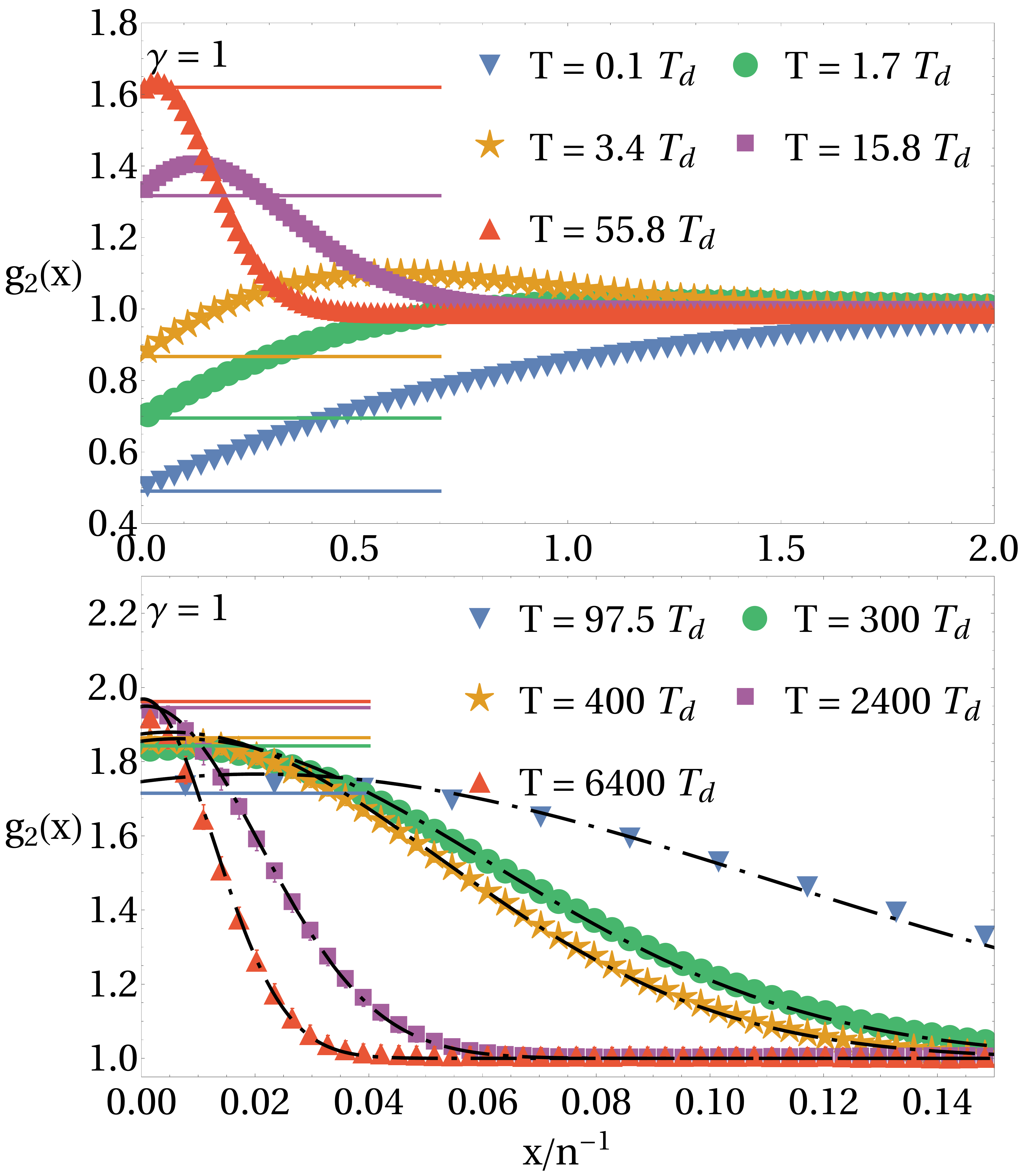}
\caption{Pair correlation function for the interaction strength $\gamma = 1$, similarly as in Fig.~\ref{fig:g201}. Symbols denote exact PIMC results. Horizontal lines represent $g_2\left(0\right)$ calculated with TBA, Eq.~\eqref{Eq:g20}, and are reported in an increasing order of the temperature from low (bottom) to high (top) values in each panel. Black dot-dashed lines show the DC limit, Eq.~\eqref{Eq:g2 DC}.
}
\label{fig:g21}
\end{figure}

\begin{figure}[h!]
\includegraphics[width=0.5\textwidth]{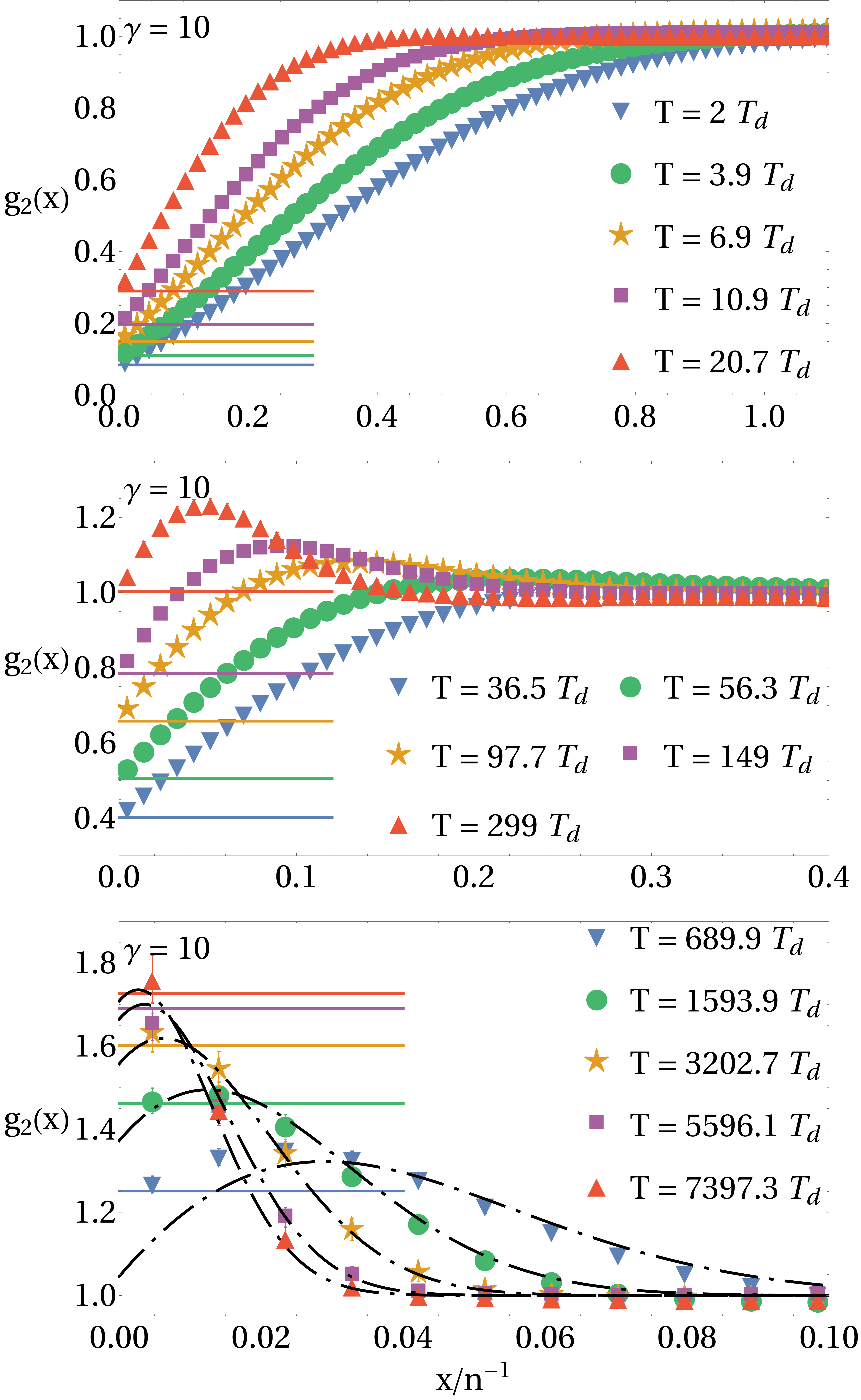}
\caption{Pair correlation function for the interaction strength $\gamma = 10$, similarly as in Fig.~\ref{fig:g201}. Symbols denote exact PIMC results. Horizontal lines represent $g_2\left(0\right)$ calculated with TBA, Eq.~\eqref{Eq:g20}, and are reported in an increasing order of the temperature from low (bottom) to high (top) values in each panel. Black dot-dashed lines show the DC limit, Eq.~\eqref{Eq:g2 DC}.
}
\label{fig:g210}
\end{figure}

In Figs.~\ref{fig:g201}-\ref{fig:g210}, PIMC results for the pair correlation function $g_2\left(x\right)$ are reported with symbols for several values of the interaction strength $\gamma$ and temperature in units of the quantum degeneracy value $T_d = \hbar^2 n^2/\left(2 m k_B\right)$. Horizontal lines denote the results of exact TBA values for the local pair correlation function $g_2\left(0\right)$, Eq.~\eqref{Eq:g20}, which is a monotonic increasing function with temperature at fixed $\gamma$. PIMC results are in a perfect agreement with the TBA predictions for the local correlations at $x=0$ and provide a full description for non-local $x>0$ values.

Below the quantum degeneracy temperature $T \ll T_d$ and for any interaction strength $\gamma$, the amplitude of the long-range decay of the pair correlation function $g_2(x)$ is proportional to $1/x^2$ and corresponds to the Fourier transform of the linear phonon expression for the static structure factor, $S(k) = \hbar|k|/(2mv)$.
Another main feature observed in $g_2(x)$ is the {\em antibunching} effect that is voiding the region around $x=0$ as the repulsive atoms tend to avoid each other.
In the limit of strong repulsion (large $\gamma$), pair correlation function approaches that of an ideal Fermi gas (IFG), $g_2(x)_{\rm IFG} \approx 1 - \sin^2(\pi n x)/(\pi n x)^2$, for which the voiding is complete, $g_2(0)_{\rm IFG}=0$ \cite{Korepin1993, Astrakharchik2006, Cherny2006, Sykes2008, Deuar2009}. Here, $g_2(x)_{\rm IFG}$ features Friedel oscillations manifested as a series of maxima located at the multiples of the mean interparticle distance $\sim 1/n$. Friedel oscillations can be interpreted as a interference effect between incident and reflected wave in the two-body collisions on an impenetrable potential. 
Friedel oscillations are also found in the density profile of a 1D interacting electron gas with an impurity \cite{Friedel1958}. Since local maxima in the pair correlation function imply the existence of more likely separations between atoms, Friedel oscillations can be also understood as a quasicrystalline order (with a period $\sim 1/n$) in the two-particle sector of the many-body wavefunction even though the density of the gas is uniform. The ideal Fermi gas limit  $g_2(x)_{\rm IFG}$ is equally valid for strong attractive interactions, i.e., for large and negative $\gamma$, by describing the pair correlations of a metastable state known as super Tonks-Girardeau gas \cite{Astrakharchik2005, Batchelor2005, Haller2009}.  
For weaker interactions (small and intermediate values of $\gamma$), $g_2(x)$ is a monotonically increasing function at very low temperatures.

Thermal fluctuations lead to a qualitatively different behavior due to the {\em bunching} effect in which the probability of two particles to overlap is enhanced, so that the local value for $g_2(x)$ increases with temperature until the maximum value $g_2(0)=2$ of the classical gas is reached.
As a result of the competition between the thermal bunching and antibunching due to the interparticle repulsion, the pair correlation function shows a global maximum larger than unity $g_2\left(x_{\rm max}\right) > 1$ located at a finite interparticle distance $x_{\rm max} > 0$. This global maximum is formed above a threshold temperature which is higher for larger values of $\gamma$.
The value in the maximum, $g_2\left(x_{\rm max}\right)$, increases while its position $x_{\rm max}$ decreases with temperature.

At high temperatures, all pair correlation functions, calculated at different values of $\gamma$, approach the Gaussian shape as predicted by the decoherent classical theory and given by Eq.~\eqref{Eq:g2 DC}. We find an excellent agreement between the PIMC results and the DC theory. In this regime, the local pair correlation function is always larger than unity, $g_2\left(0\right) > 1$. The maximum value is achieved for local correlations $x\to 0$, where strongest bunching effect is observed. The limiting value $g_2\left(0\right) \to 2$ is the same as in an ideal Bose gas and is fully driven by the thermal energy. The DC regime is achieved at higher temperatures for stronger interactions \cite{DeRosi2021II} and the value of the global maximum in the pair correlation function is approximated by $g_2\left(x_{\rm max}\right)_{\rm DC} = 2 - \gamma \sqrt{2 \pi/\tau} + 2 \gamma^2/\tau $ which is located at $x_{\rm max} = 2 \gamma/\left( n  \tau \right) \ll n^{-1}$ \cite{Sykes2008, Deuar2009} where $\tau = T/T_d$. 
This maximum is more clearly visible by increasing $\gamma$, Figs.~\ref{fig:g201}-\ref{fig:g210}.

In Fig.~\ref{fig:g201}, we show the characteristic examples of the pair correlation function in the weakly-interacting Gross-Pitaevskii regime with $\gamma = 10^{-1}$. 
Each panel corresponds to different thermal regimes in order of increasing temperature values: hole anomaly (upper), intermediate (middle) and DC regime (lower) \cite{DeRosi2021II}. In the limit of zero temperature, $g_2\left(0\right) \approx 0.7$ for this value of $\gamma$, that there is a weak suppression of the probability of two particles to overlap, as compared to the uncorrelated (ideal Bose gas) value, $g_2(0) = 1$. At very low temperature, there is a {\em quasicondensate} in the system, which is characterized by suppressed density fluctuations due to repulsive interactions \cite{Esteve2006} and enhanced long-wavelength phase fluctuations due to thermal excitations \cite{Dettmer2001}. The resulting local pair correlation function indicates second-order coherence $g_2\left(0\right) \approx 1$, for which there exists a finite probability that two particles come close to each other. So that, the existence of the quasicondensate is fully driven by the correlations induced by very weak interatomic repulsion, similarly to the case of the DC regime $g_2\left(0\right) \approx 2$ where interaction effects can be neglected.

In Fig.~\ref{fig:g21}, we report the pair correlation function for the intermediate value of the interaction strength $\gamma = 1$ in the hole anomaly (upper panel) and in the DC (lowest panel) regimes \cite{DeRosi2021II}. This is the most difficult regime to describe as it cannot be treated perturbatively and one has to rely on numerics in order to obtain accurate results. At zero temperature, $g_2\left(0\right) \approx 0.5$, in agreement with thermal Bethe-Ansatz calculations. The potential energy of two-body interactions, $E_{\rm pot}/N = g_2(0)  g n /2$, is maximal in this regime, as compared to the Gross-Pitaevskii regime with vanishing coupling constant, $g\to 0$, and the Tonks-Girardeau regime of impenetrable particles, $g_2(0)=0$.
As the temperature is increased, antibunching effect becomes less prominent, until temperature $T\approx 4T_d$ is reached with $g_2\left(0\right) \approx 1$.
Further increase in the temperature leads to bunching effect.

In Fig.~\ref{fig:g210}, we show the pair correlation function in the case of strong repulsion, $\gamma = 10$.
Different thermal regimes are reported in each panel: hole anomaly (upper), virial hard-core (HC) and beyond (middle) and DC regime (lower) \cite{DeRosi2021II}. At zero temperature, the local pair correlation function approaches zero $g_2\left(0\right) \approx 0.1$, by signalling the antibunching effect. The gas is then fermionized as the strong repulsion mimics the Pauli exclusion principle for intrinsic fermions at zero relative distance and pairs of bosons are thus never at the same spatial position, $g_2(0)=0$. The antibunching is washed out by thermal effects.
At the lowest temperature here reported $T = 2 T_d$, Friedel oscillations are not visible at larger distances, by confirming that they exist only below the quantum degeneracy point $T \ll T_d$.

In the virial hard-core regime (we refer to the middle panel of Fig.~\ref{fig:g210}) of strong interactions and high temperature $\pi^2 < \tau \lesssim \gamma^2$ \cite{DeRosi2019, DeRosi2021II} the local pair correlation function does not manifest complete antibunching $g_2\left(0\right) = 0$, as witnessed by exact TBA and PIMC results.
Instead, its value remains finite ($0.4 \lesssim g_2\left(0\right) \lesssim 0.6$) which is in contrast to what is claimed in Refs.~\cite{Sykes2008, Deuar2009}.

The crossover from the virial HC to the DC regime can be induced by increasing the temperature~\cite{DeRosi2021II}. It is signalled by the global maximum in the pair correlation function emerging at temperature $\tau \approx \gamma^2$ corresponding to $\sigma \sim |a|$~\cite{Deuar2009} where $a$ is the 1D scattering length and $\sigma$, up to some multiplicative constant, corresponds to the thermal de Broglie wavelength. Such a crossover is explained by the interplay between the thermal bunching and repulsion antibunching effects acting on comparable scales $\sigma \sim |a|$ \cite{Deuar2009}. In the crossover regime $\gamma^2/\tau \simeq 0.1-0.4$ (middle and lowest panels of Fig.~\ref{fig:g210}), the local pair correlation function exhibits a behavior similar to the quasicondensate $g_2\left(0\right) \approx 1$, indicating local second-order coherence. However, unlike the quasicondensate regime, the non-local correlations on distance scale $|x| \sim \sigma \sim |a|$ are not coherent but bunched as witnessed by the presence of the maximum \cite{Deuar2009}.

In the high temperature regime (lowest panel of Fig.~\ref{fig:g210}), we find an excellent agreement between the PIMC results and the analytical limit provided by the decoherent classical theory, Eq.~\eqref{Eq:g2 DC}, except for the lowest value of temperature for which the deep DC regime is not yet achieved \cite{Kheruntsyan2003, Sykes2008, Deuar2009, DeRosi2021II}.

It is important to point out that other known analytical limits for the pair correlation function $g_2\left(x\right)$ (weakly- and strongly-interacting at high temperatures and decoherent quantum regimes) \cite{Sykes2008, Deuar2009} are not reported in the present work as they do not show agreement with our exact PIMC results. Comparison with the approximate theories of the weakly- and strongly-interacting quantum regimes \cite{Sykes2008, Deuar2009}
is not possible due to the growing computational cost of PIMC method going down to extremely low temperatures. 
While our exact PIMC and TBA results for the local pair correlation function $g_2\left(0\right)$ agree for any considered value of the interaction strength and temperature, they do not match with the corresponding analytical limits \cite{Kheruntsyan2003} in some regimes. Given the proportionality with Tan's contact parameter, Eq.~\eqref{Eq:g20}, $g_2\left(0\right)$ is an additional thermodynamic property. New analytical descriptions have been built and they exhibit an excellent agreement with TBA in their regimes of validity and for several thermodynamic quantities \cite{DeRosi2017, DeRosi2019, DeRosi2021II}.

\section{Static Structure Factor}
\label{Sec:S(k)}

The static structure factor $S\left(k \right)$ is directly related to the Fourier transform of the pair correlation function, Eq.~\eqref{Eq:g2} \cite{Pitaevskii2016}:
\begin{equation}
\label{Eq:S def}
S\left( k  \right) = 1 + n \int_{-\infty}^{+\infty} dx  \left[ g_2(x) -1   \right] \cos\left(k \cdot x\right),
\end{equation}
and quantifies the spatial correlations in momentum space, where $k = k_1 - k_2$ is the relative momentum of atoms.

At small $k$, the static structure factor is sensitive to thermal and dynamical correlations and at $k = 0$ is related to the isothermal compressibility $\kappa_T$, Eq.~\eqref{Eq:kTINV} \cite{Pitaevskii2016}
\begin{equation}
\label{Eq:S(0)}
S\left(0\right) = k_B T \kappa_T .
\end{equation}
Its value can be computed exactly through the thermal Bethe-Ansatz calculation of $\kappa_T $ \cite{Yang1969, Yang1970, Panfil2014, DeRosi2021II} which has been recently compared with corresponding PIMC results, by showing an excellent agreement~\cite{DeRosi2021II}.
Equation~\eqref{Eq:S(0)} is a general result following from the fluctuation-dissipation theorem and holds for any value of temperature and interaction strength.

For large momenta, $k\to \infty$, the static structure factor approaches the uncorrelated value, $S \left(|k| \gg n \right) \to 1$ \cite{Pitaevskii2016}. 

If the excitation spectrum $\epsilon(k)$ is exhausted by a coherent single-mode (SM) quasiparticle \cite{DeRosi2021II} then the static structure factor can be approximated as follows \cite{Pitaevskii2016}
\begin{equation}
\label{Eq:S spectrum}
S_{\rm SM}\left( k \right) = \frac{\hbar^2 k^2}{2 m \epsilon\left(k\right)} \coth\left[ \frac{\epsilon\left(k\right)}{2 k_B T}\right] .
\end{equation}
The prefactor in Eq.~\eqref{Eq:S spectrum} recovers the Feynman relation~\cite{Feynman54} and corresponds to the static structure factor at zero temperature $\hbar^2 k^2/\left[2 m \epsilon\left(k\right)\right]$. In a previous publication \cite{DeRosi2021II}, PIMC results for $S\left(k\right)$ were used to estimate the spectrum within the Feynman approximation and to show that it is valid for temperatures below the hole anomaly value. The low-temperature correction in Eq.~\eqref{Eq:S spectrum} is derived from the principle of the detailed balance.

By using the $T = 0$ Bogoliubov (BG) spectrum $\epsilon\left(k\right) = \sqrt{\left( \hbar k v\right )^2 + \left( \hbar^2 k^2/2 m\right)^2}$ \cite{Lieb1963, Lieb1963II} in Eq.~\eqref{Eq:S spectrum}, one can analytically obtain the static structure factor at low temperature in the weakly-interacting regime $\tau \ll \sqrt{\gamma} \ll 1$ \cite{Kheruntsyan2003, Sykes2008, Deuar2009, DeRosi2019, DeRosi2021II}.
For large momenta, one can use the expansion of the BG spectrum $\epsilon(\hbar |k|  \gg m v) \sim \hbar^2 k^2/(2 m) + m v^2$ in Eq.~\eqref{Eq:S spectrum} and obtain at the level of the Feynman approximation \cite{Pitaevskii2016}:
\begin{equation}
\label{Eq:S large k}
S_{\rm BG}\left( \hbar |k|  \gg m v\right) = 1 - \frac{2 m^2 v^2}{\hbar^2 k^2} \ ,
\end{equation}
which recovers the model-independent limit $S\left(|k| \gg n\right) \to 1$ after a further expansion $\hbar |k| \gg m v$. For arbitrary values of the interaction strength $\gamma$, the first term in Eq.~(\ref{Eq:S large k}) is recovered by assuming the free-particle excitation spectrum, $\epsilon\left(k\right) = \hbar^2k^2/(2m)$
in Eq.~\eqref{Eq:S spectrum} within the Feynman approximation. 
Equations \eqref{Eq:S spectrum} and \eqref{Eq:S large k} are very general as they are also valid for a Bose gas in three spatial dimensions \cite{Pitaevskii2016}.

At small momenta, $\hbar |k| \ll m v$, the excitation spectrum is dominated by linear phonons $\epsilon\left(k\right) = v \hbar |k|$ propagating with sound velocity $v$.
This allows to obtain from Eq.~\eqref{Eq:S spectrum} a generic expression for the static structure factor at small momenta,
\begin{equation}
\label{Eq:S phonons}
S_{\rm LL}\left( \hbar |k|  \ll m v\right) = \frac{\hbar |k|}{2 m v} \coth\left( \frac{v\hbar |k|}{2 k_B T}\right).
\end{equation}
The speed of sound itself is related to the zero-temperature compressibility,
\begin{equation}
\label{Eq:compressibility T=0}
mv^2 = \kappa_{T = 0}^{-1} = n \frac{\partial \mu_0}{\partial  n}, 
\end{equation}
and can be obtained from the equation of state, $\mu_0(n)$, which is expressed in terms of the chemical potential at zero temperature $\mu_0$. 
Equation \eqref{Eq:S phonons} is then a reliable small-$k$ approximation for the static structure factor which corresponds to the Luttinger Liquid (LL) regime holding at very low temperatures $k_B T \ll \mu_0$ and for any interaction strength $\gamma$ \cite{Haldane1981, DeRosi2017}. Eq.~\eqref{Eq:S phonons} recovers the classical result, Eq.~\eqref{Eq:S(0)}, where the isothermal compressibility is calculated at zero temperature $\kappa_{T = 0} = 1/\left(m v^2\right)$.

\begin{figure}[h!]
\includegraphics[width=0.5\textwidth]{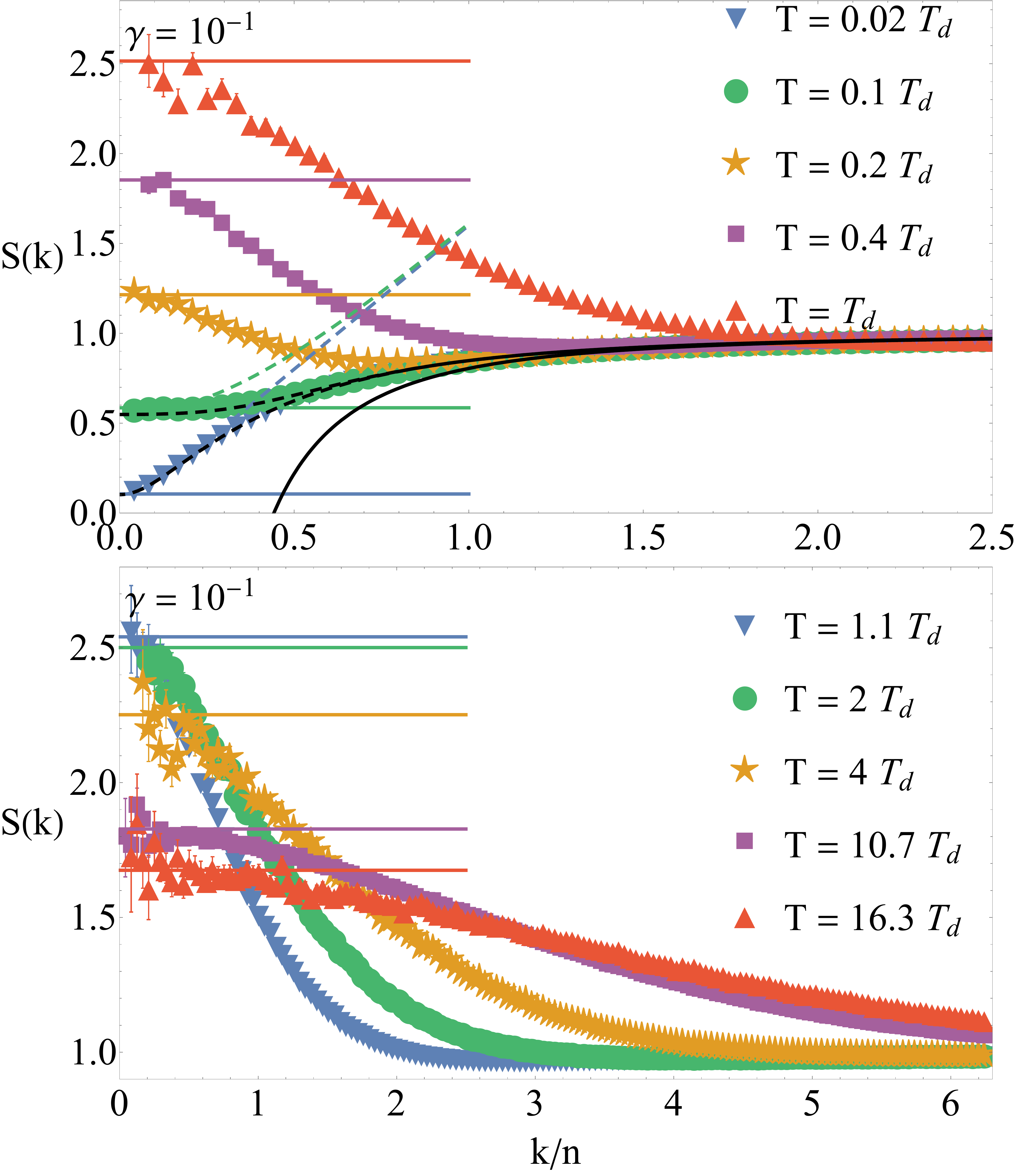}
\caption{Static structure factor $S\left(k\right)$ as a function of the relative momentum $k$ in units of the density $n$ for the interaction strength $\gamma = 10^{-1}$. Different values of temperature in units of the quantum degeneracy value $T_d = \hbar^2 n^2/\left(2 m k_B\right)$ are reported.
Symbols denote exact Path Integral Monte Carlo results. Horizontal lines represent the static structure factor at zero momentum $S\left(0\right)$ calculated with thermal Bethe Ansatz, Eq.~\eqref{Eq:S(0)}. Dashed coloured lines show Eq.~\eqref{Eq:S phonons} calculated with the phononic spectrum. Black dashed lines denote Eq.~\eqref{Eq:S spectrum} calculated with the Bogoliubov spectrum. All limits are reported in an increasing order of the temperature from low (bottom) to high (top) values in the upper panel and in the reversed order in the lower panel.
Black solid curve represents the large-$k$ behavior of the Bogoliubov approximation, Eq.~\eqref{Eq:S large k},  which is independent of temperature.
}
\label{fig:Sk01}
\end{figure}

\begin{figure}[h!]
\includegraphics[width=0.5\textwidth]{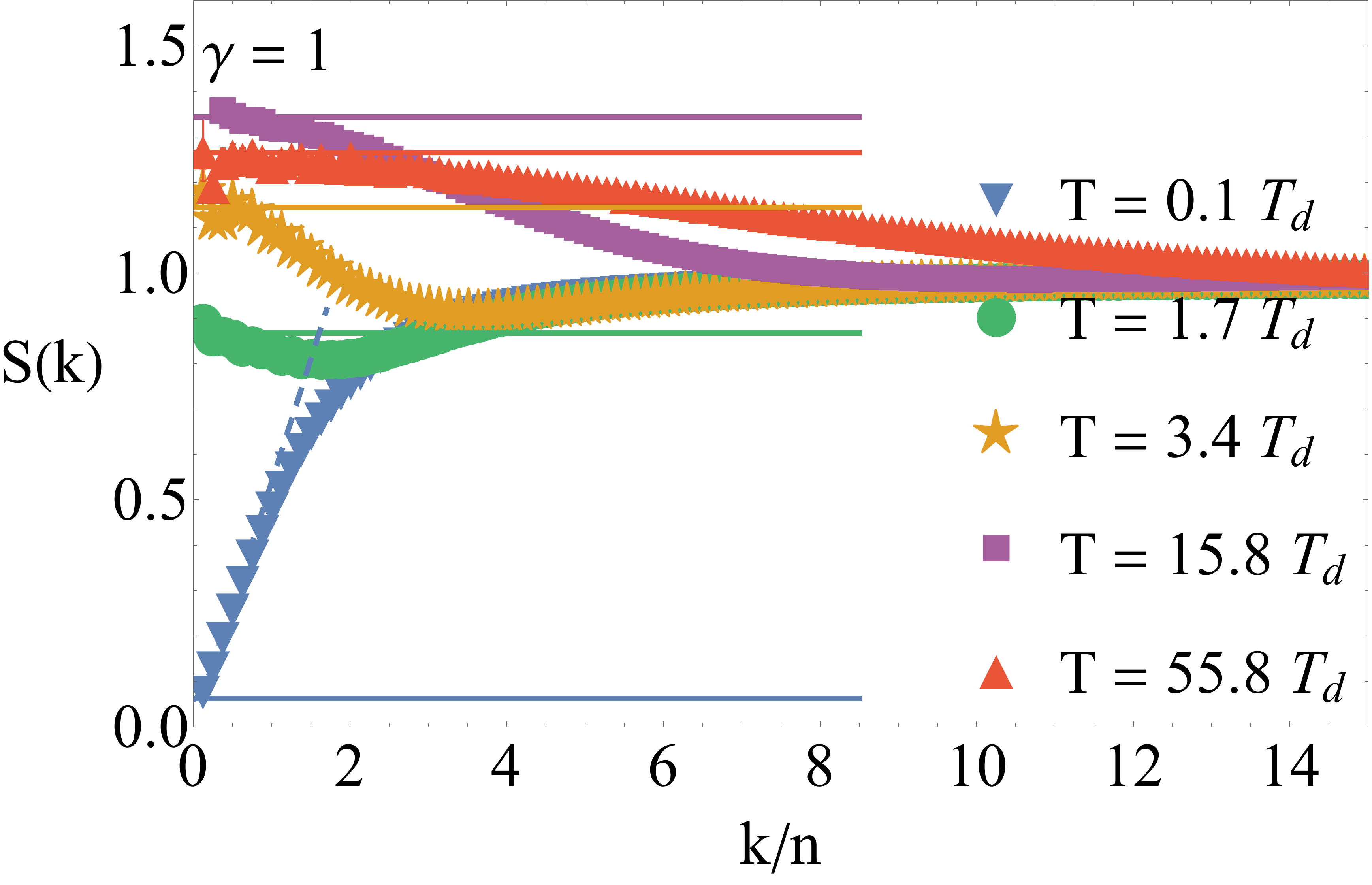}
\caption{Static structure factor for the interaction strength $\gamma = 1$, similarly as in Fig.~\ref{fig:Sk01}. Symbols denote exact PIMC results. Horizontal lines represent $S\left(0\right)$ calculated with TBA, Eq.~\eqref{Eq:S(0)}. Dashed line shows Eq.~\eqref{Eq:S phonons} calculated with the phononic spectrum.
}
\label{fig:Sk1}
\end{figure}

\begin{figure}[h!]
\includegraphics[width=0.5\textwidth]{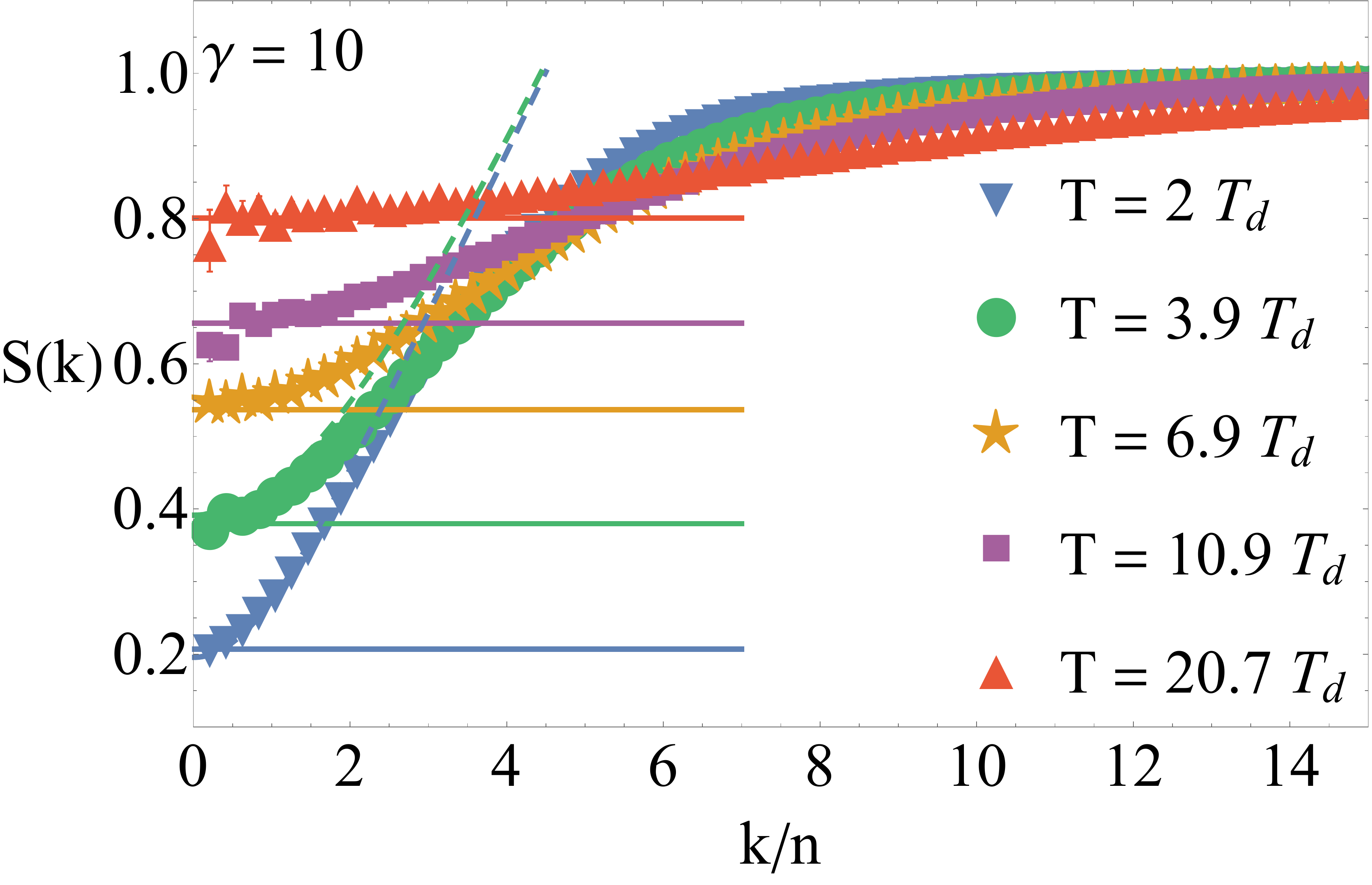}
\caption{
Static structure factor for the interaction strength $\gamma = 10$, similarly as in Fig.~\ref{fig:Sk01}. Symbols denote exact PIMC results. Horizontal lines represent $S\left(0\right)$ calculated with TBA, Eq.~\eqref{Eq:S(0)}. Dashed lines show Eq.~\eqref{Eq:S phonons} calculated with the phononic spectrum. All the lines are reported in an increasing order of temperature from low (bottom) to high (top) values.
}
\label{fig:Sk10}
\end{figure}

In Figs.~\ref{fig:Sk01}-\ref{fig:Sk10}, we report the static structure factor $S\left(k\right)$ calculated for characteristic values of the interaction strength $\gamma$ and temperature in units of the quantum degeneracy value $T_d = \hbar^2 n^2/\left(2 m k_B\right)$.
In Fig.~\ref{fig:Sk01}, we report $S(k)$ in the weakly-interacting GP regime with $\gamma = 10^{-1}$. Each panel corresponds to a different physical thermal regime ordered by increasing temperature values: hole anomaly (upper) and intermediate (lower) \cite{DeRosi2021II}.
In the upper panel, black dashed lines denote Eq.~\eqref{Eq:S spectrum} calculated with the Bogoliubov spectrum and black solid line represents the corresponding large-$k$ behavior, Eq.~\eqref{Eq:S large k}.
We show $S(k)$ calculated in the regime of the intermediate interactions, $\gamma=1$, in Fig.~\ref{fig:Sk1}. 
The case of strong correlations, $\gamma=10$, is considered in Fig.~\ref{fig:Sk10}.

Horizontal lines denote the exact zero-momentum static structure factor $S\left(0\right)$ as calculated by TBA, Eq.~\eqref{Eq:S(0)}.
In the limit of zero temperature, zero value is recovered, $S\left(0\right) = 0$, for any $\gamma$.
$S\left(0\right)$ quantifies the interplay between the quantum and thermal fluctuations.
The effect of the temperature is more prominent in weakly-interacting systems.
For example, at the degeneracy temperature $T=T_d$, the thermal effects are large for weak interactions, $\gamma=10^{-1}$, and moderate for strong interactions, $\gamma=10$.
In fact, in the weakly-repulsive regime, the inverse isothermal compressibility $\kappa_T^{-1}$, Eq.~\eqref{Eq:kTINV}, is small and $S\left(0\right)$ grows rapidly as the temperature is increased. In the opposite fermionized regime with $\gamma \gg 1$, $\kappa_T^{-1}$ is large and $S\left(0\right)$ remains small even at relatively high temperature.
A characteristic thermal crossover temperature, defined by the condition $S(0)=1$, is an increasing function of $\gamma$ and reaches values as large as $\tau = T/T_d \sim 10^2$ (not reported in Fig.~\ref{fig:Sk10}). Above this temperature threshold, the static structure factor develops a global maximum $S(k_{\rm max}) = S(0) > 1$ which defines the most likely momentum $k_{\rm max}$ of atoms. 
The zero-momentum value, $S\left(0\right)$, is a non-monotonic function of temperature and it exhibits a maximum occurring at a certain temperature $\tau_{\rm max}$.
The temperature $\tau_{\rm max}$ itself increases monotonically with increasing the strength of interactions $\gamma$.
This temperature is higher than that of the hole anomaly and smaller than that of the DC regime \cite{DeRosi2021II} for any interaction strength. Such universal result holds even for the largest considered interaction strength, $\gamma = 10$, where the maximum of $S\left(0\right)$ occurs at temperature around $\tau_{\rm max} \sim10^4$ not shown in Fig.~\ref{fig:Sk10}.

It is instructive to identify the temperatures and momenta for which the Luttinger Liquid description, Eq.~\eqref{Eq:S phonons}, is applicable. 
This theory is based on assuming a linear phonon dispersion relation and has the same underlying assumption as the single-mode result given by Eq.~\eqref{Eq:S spectrum}. By making a comparison of this effective Luttinger Liquid theory (coloured dashed lines in Figs.~\ref{fig:Sk01}-\ref{fig:Sk10}) with the exact numerical PIMC result, one finds that the single-mode phonon assumption holds well at low momenta and low temperatures.
For a fixed value of $\gamma$, such low-$k$ approximation holds for a broader range of $k$ at lower temperatures.
The range of applicability gets broader by increasing $\gamma$ provided the Luttinger-Liquid condition $\hbar |k| \ll m v$ being $v$ a monotonic increasing function with $\gamma$ \cite{Lieb1963II, DeRosi2017}. 
Our results also confirm the temperature-range of validity of the Luttinger-Liquid description $k_B T \ll \mu_0$. Such a range gets broader by increasing the interaction strength as the zero-temperature chemical potential $\mu_0$ is an increasing function with $\gamma$ \cite{Lieb1963, DeRosi2017}. 

All analytical limits agree in their regimes of validity with the corresponding PIMC results calculated at the same value of temperature and for any interaction strength.

For each set of results at fixed $\gamma$, the decoherent classical regime at high temperature is not reported as the static structure factor is trivially a constant function for any value of $k$ and approaches the model-independent value $S\left(k\right) \approx 1$.

\section{One-body density matrix}
\label{Sec:OBDM}

The one-body density matrix (OBDM) quantifies the coherence and is defined as \cite{Pitaevskii2016}
 \begin{equation}
 \label{Eq:OBDM}
 g_1\left (x = x_1 - x_2 \right) = \langle  \hat{\psi}^\dagger\left(x_1\right)  \hat{\psi}\left(x_2\right) \rangle.
 \end{equation}
It is proportional to the amplitude of the process in which one particle is annihilated at position $x_2$ and the same state is recovered by creating a particle at position $x_1$.
Definition~\eqref{Eq:OBDM} applies to any value of the interaction strength and temperature.
For a zero displacement, $x = 0$, one recovers the diagonal density of the system which, in the uniform configuration studied here, exactly provides the linear density $n$. The momentum distribution $n\left(k\right)$ which is a function of the momentum $k$, is the Fourier transform of the OBDM \cite{Pitaevskii2016}.

At very high temperatures, $T \gg T_d$, the system behaves as a classical gas and follows the Maxwell-Boltzmann (MB) statistics. The OBDM exhibits the Gaussian behavior
\begin{equation}
\label{Eq:g1 gaussian}
g_1\left(x \right)_{\rm MB}  = n e^{-x^2/(2\sigma^2)} \ ,
\end{equation}
and vanishes at distances much larger than the standard deviation $\sigma = \lambda/\sqrt{2 \pi}$ which is proportional to the thermal wavelength $\lambda$, already defined in Eq.~\eqref{Eq:g2 DC}.

At small distance $x$, the OBDM can be expanded as a sum of analytic and non-analytic terms~\cite{Astrakharchik2006II} and holding for any value of the interaction strength and temperature:
\begin{equation}
\label{Eq:OBDM small x}
\frac{g_1\left(x \to 0\right)}{n} = 1 + \sum_{i = 1}^\infty c_i \left(n x\right)^{i} + b_3 \left|n x\right|^3 + \mathcal{O}\left(|n x|^4\right).
\end{equation}
The coefficients $c_i$ of the Taylor expansion of the analytic part of the OBDM are the corresponding moments of the momentum distribution $n\left(k\right)$ \cite{Astrakharchik2006II}, they diverge for $i > 3$ and the odd coefficients vanish $c_1 = c_3 = \cdots = 0$ due to reflection symmetry. From the Hellmann-Feynman theorem \cite{Feynman1939} one finds:
\begin{equation}
\label{Eq:c2}
c_2 = - \frac{1}{2} \left( \frac{E}{N} \frac{2 m}{\hbar^2 n^2} - \frac{\mathcal{C}}{N }  \frac{1}{\gamma n^3}  \right) .
\end{equation}
It is easy to show that the second coefficient, Eq.~\eqref{Eq:c2}, can be rewritten in terms of the average kinetic energy $\langle H_{\rm kin}\rangle$ \cite{Pitaevskii2016}, see Appendix \ref{Sec:c2}.
The non-analytic part of the OBDM expansion starts as $|n x|^3$ with the coefficient $b_3$ which is related to the high-momentum tail of $n\left(k\right)$, as it depends on the Tan's contact $\mathcal{C}$
\begin{equation}
\label{Eq:c3}
b_3 = \frac{\mathcal{C}}{N} \frac{1}{12 n^3} \ .
\end{equation}
Both the internal energy and Tan's contact per particle, $E/N$ and $\mathcal{C}/N$ respectively, in Eqs.~\eqref{Eq:c2}-\eqref{Eq:c3} have been evaluated exactly with thermal Bethe Ansatz \cite{Yang1969, Yang1970, DeRosi2019, DeRosi2021II}. Equations~\eqref{Eq:OBDM small x}-\eqref{Eq:c3} for the short-range expansion of the OBDM have been derived at $T = 0$ \cite{Olshanii2003} and here they are generalized to a finite temperature whose dependence enters in $E$ and $\mathcal{C}$.

In higher spatial dimensions, where the Bose-Einstein condensate emerges below the critical temperature, the presence of Off-Diagonal Long-Range Order in homogeneous systems is manifested by a finite value of the OBDM at large distances, which corresponds to the condensate fraction \cite{Pitaevskii2016}.
In our system instead, the OBDM always vanishes at large distances, $g_1 \to 0$, i.e., there is no Bose-Einstein condensation even at zero temperature.

Within the Luttinger Liquid theory, applicable at very low temperature $k_B T  \ll \mu_0$ and for any interaction strength $\gamma$, the OBDM exhibits a power-law decay \cite{Schwartz1977, Haldane1981, Cazalilla2004, Pitaevskii2016}:
\begin{equation}
\label{Eq:g1 power}
\frac{g_1\left(\xi \ll |x| \ll x_T\right)_{\rm LL}}{n} = \left(  \frac{K}{\pi n |x|} \right)^{\frac{1}{2 K}}
\end{equation}
holding for distances much larger than the healing length $\xi = \hbar/(\sqrt{2} m v)$ and smaller than the thermal correlation radius $x_T = \hbar v/\left( k_B T \right)$. The Luttinger Liquid parameter $K = v_F/v$ can be calculated exactly with Bethe Ansatz for any interaction strength $\gamma$, via the corresponding results of the speed of sound $v$.
The power-law decay~\eqref{Eq:g1 power} has been shown to be accurate for any value of $K$ by comparison with Diffusion Monte Carlo results at zero temperature \cite{Cazalilla2004}, where quantum fluctuations depend explicitly on the value of $v$ \cite{Pitaevskii2016}.
The vanishing asymptotic value of the one-body density matrix at large distances excludes the existence of Bose-Einstein condensation in infinite systems \cite{Schultz1963}. However, due to the smallness of the exponent $1/(2 K)$ in the weakly-interacting GP regime, $\gamma \ll 1$, the OBDM does not vanish at macroscopic distances $|x| \gg \xi$,  signalling the presence of a quasicondensate exhibiting features of superfluids \cite{Astrakharchik2004}.

\begin{figure}[h!]
\includegraphics[width=0.5\textwidth]{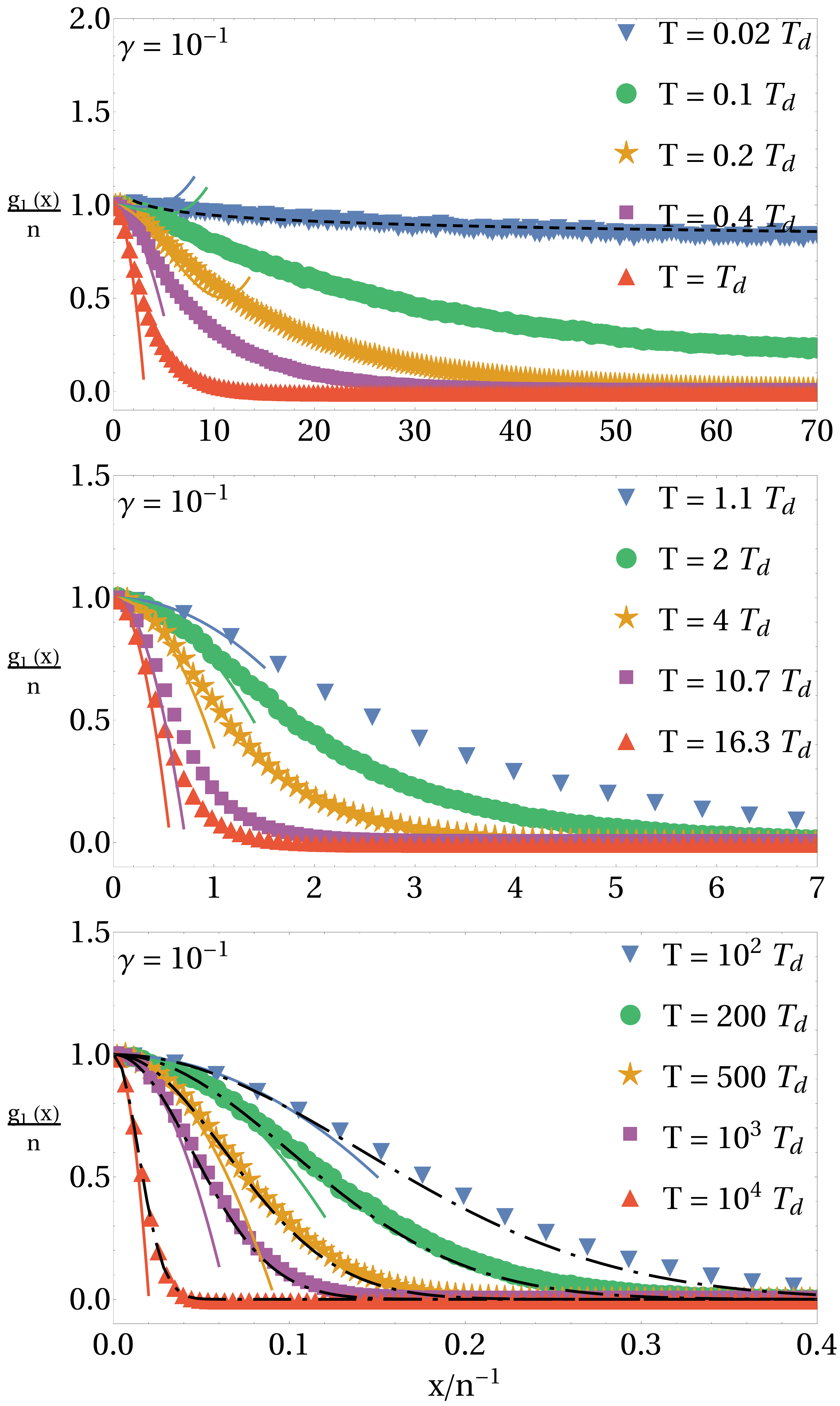}
\caption{One-body density matrix $g_1\left(x\right)/n$ as a function of distance $x$ in units of the mean interparticle separation $n^{-1}$ for the interaction strength $\gamma = 10^{-1}$. Different values of temperature in units of the quantum degeneracy value $T_d = \hbar^2 n^2/\left(2 m k_B\right)$ are reported.
Symbols denote exact Path Integral Monte Carlo results. Symbol size is larger than the statistical error bar. Solid lines represent the expansion at small distances, Eq.~\eqref{Eq:OBDM small x}, calculated with thermal Bethe Ansatz. In the upper panel, black dashed line denotes the Luttinger-Liquid power-law decay, Eq.~\eqref{Eq:g1 power}, which is independent of temperature. In the lower panel, black dot-dashed lines correspond to the Maxwell-Boltzmann behavior, Eq.~\eqref{Eq:g1 gaussian}. All curves are reported in an increasing order of the temperature from low (top) to high (bottom) values in each panel.
}
\label{fig:g101}
\end{figure}

\begin{figure}[h!]
\includegraphics[width=0.5\textwidth]{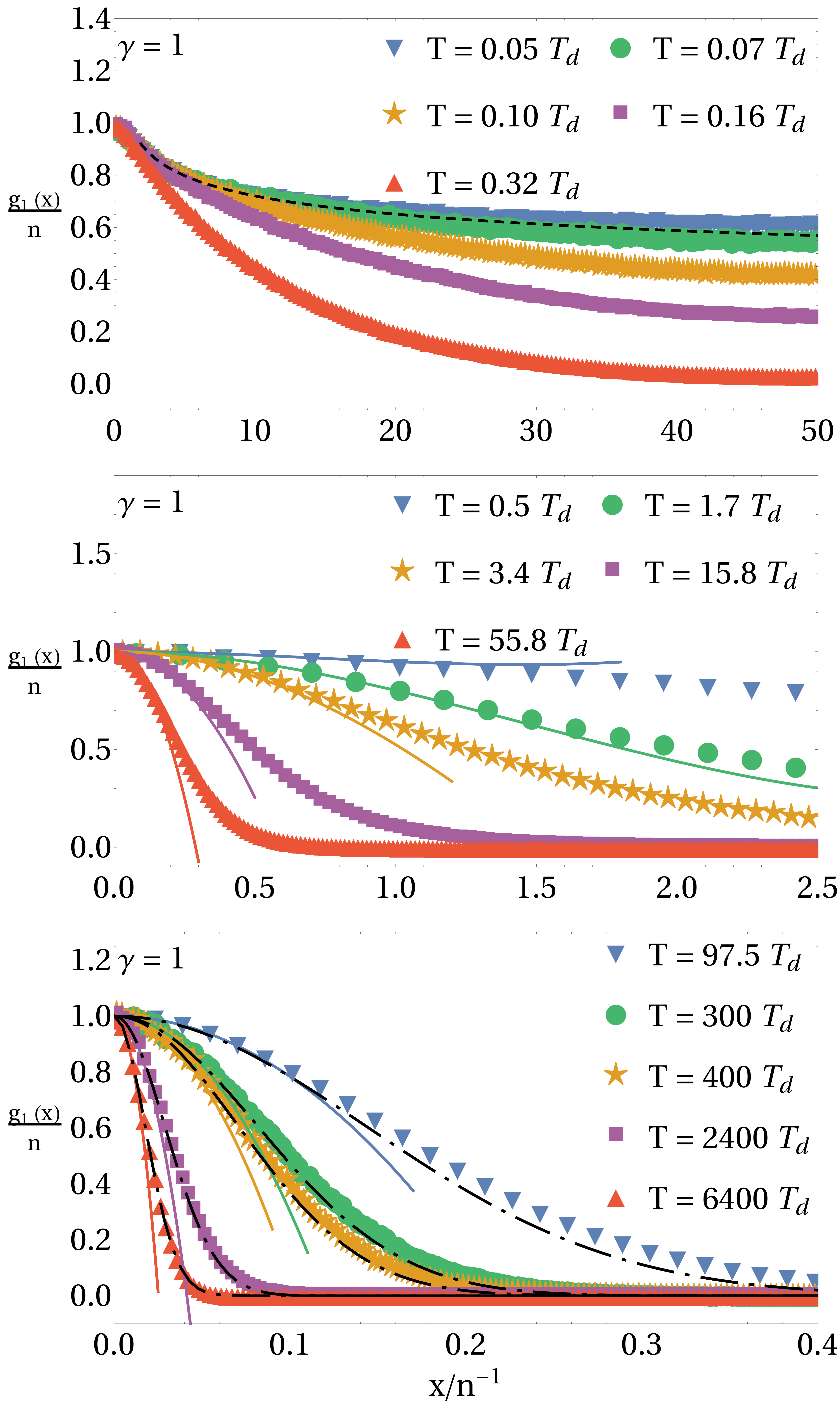}
\caption{OBDM for the interaction strength $\gamma = 1$, similarly as in Fig.~\ref{fig:g101}. Symbols denote exact PIMC results. Solid lines represent the expansion at small distances, calculated with TBA, Eq.~\eqref{Eq:OBDM small x}. Black dashed line denotes the LL power-law decay, Eq.~\eqref{Eq:g1 power}, while black dot-dashed curves correspond to the MB behavior, Eq.~\eqref{Eq:g1 gaussian}. All the lines are reported in an increasing order of temperature from low (top) to high (bottom) values in each panel.
}
\label{fig:g11}
\end{figure}

\begin{figure}[h!]
\includegraphics[width=0.5\textwidth]{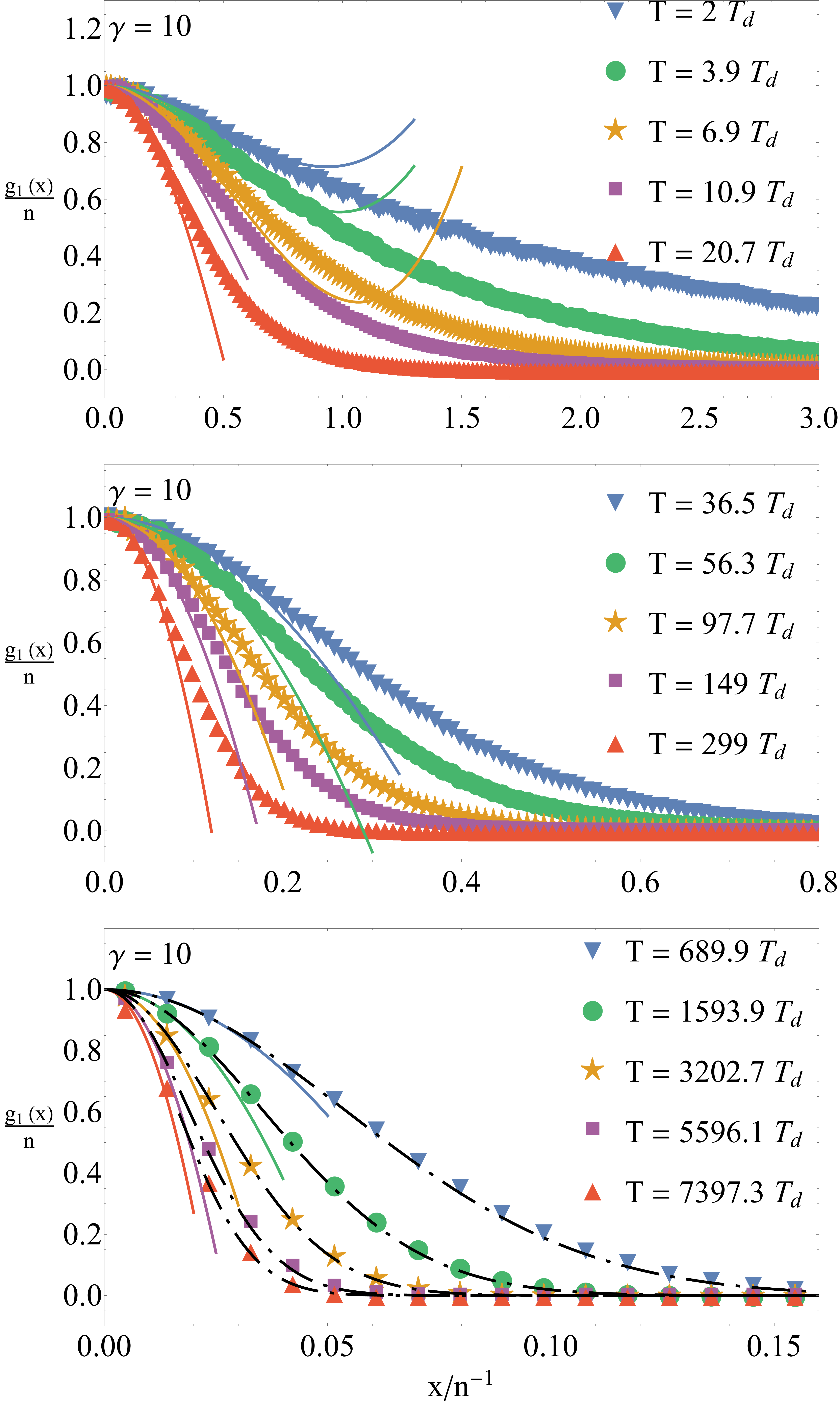}
\caption{OBDM for the interaction strength $\gamma = 10$, similarly as in Fig.~\ref{fig:g101}. Symbols denote exact PIMC results. Solid lines represent the expansion at small distances, calculated with TBA, Eq.~\eqref{Eq:OBDM small x}. Black dot-dashed curves correspond to the MB behavior, Eq.~\eqref{Eq:g1 gaussian}. All the lines are reported in an increasing order of temperature from low (top) to high (bottom) values in each panel.
}
\label{fig:g110}
\end{figure}

In Figs.~\ref{fig:g101}-\ref{fig:g110}, we show with symbols the PIMC results of the one-body density matrix $g_1\left(x\right)$ for different values of the interaction strength $\gamma$ and temperature. Solid lines denote the short-range expansion, Eq.~\eqref{Eq:OBDM small x}, calculated from TBA quantities. In the upper panel of the sets of results for $\gamma = 10^{-1}$ and~$1$, the black dashed line denotes the Luttinger-Liquid quasi long-range decay, Eq.~\eqref{Eq:g1 power}, valid in the limit of zero temperature. Instead, the power-law decay~\eqref{Eq:g1 power} is not reported for $\gamma = 10$ as in this strongly-interacting regime only temperatures beyond the validity of the LL theory have been explored.
The DC regime of high temperatures reported in the lower panels of Figs.~\ref{fig:g101}-\ref{fig:g110} allows a comparison with the
Gaussian behavior expected for a classical gas, Eq.~\eqref{Eq:g1 gaussian}, and the corresponding predictions are shown with black dot-dashed lines. We notice that for any interaction strength, a better agreement with PIMC results is found at small distances by pointing out that, in the DC regime, quantum statistics still play a role at large interatomic separation despite the high temperature.
All analytical limits show a fair agreement with the corresponding PIMC results calculated at the same value of temperature.

In Fig.~\ref{fig:g101}, we show the OBDM in the weakly-interacting GP regime with $\gamma = 10^{-1}$. Each panel corresponds to different physical thermal regimes in order of increasing temperature values: hole anomaly (upper), intermediate (middle) and DC regime (lower) \cite{DeRosi2021II}. It is worth noticing that although the OBDM vanishes at large distances, in agreement with the absence of the Bose-Einstein condensation in one dimension, still the OBDM can be significantly different from zero at low temperatures, even at distances large compared to the mean interparticle distance $x > n^{-1}$. In the considered example, the coherence is preserved at distances of the order of hundreds mean interparticle distances. Such a behavior is often referenced as a quasicondensation and it is responsible for making the Gross-Pitaevskii and Bogoliubov theories still applicable in 1D. Instead, as the temperature is increased, thermal fluctuations significantly suppress the coherence, making it disappear for $x\sim n^{-1}$ (middle panel) and even smaller distances (lower panel). In the latter case, the comparison with the Maxwell-Boltzmann prediction (black dot-dashed lines in the lower panel) makes it clear that at high temperatures the coherence is lost at distances of the order of $\sigma$.

The regime of the intermediate interaction strength, $\gamma=1$, is the most difficult one for an analytic description as there is no separation of scales and the scattering length is of the same order as the mean interparticle distance $a \sim n^{-1}$.
Furthermore, at low temperature, the kinetic energy is of the same order as the potential interaction energy.
The OBDM in this interaction regime is shown in Fig.~\ref{fig:g11}.
We observe that at low temperatures the OBDM is significantly different from zero up to distances of the order of tens of the interparticle distances, making the quasicondensate concept applicable even to this case.
The coherence is instead lost when the temperature becomes comparable with the rest of the equal energy scales, i.e. for $k_B T \sim k_B T_d = \hbar^2 n^2/\left(2 m\right) \sim \hbar^2/(ma^2)$.
Thus, $\tau = T/T_d = 1$, separates the quantum low-temperature and decoherent classical regimes. In the upper panel of Fig.~\ref{fig:g11}, we report results at very low temperature and large distances where the short-distance expansion, Eq.~\eqref{Eq:OBDM small x}, is not shown.

The regime of large values of the interaction strength $\gamma$ is very interesting as the Girardeau's mapping applies here, that is the wavefunction of bosons with infinite $\gamma$ equals the absolute value of the wavefunction of the ideal fermions in 1D \cite{Girardeau1960}. From this mapping, diagonal properties (pair correlation function, static structure factor, etc.) and the energy are the same in the two systems. Instead, the momentum distribution and the OBDM are manifestly different. Indeed, the momentum distribution of an ideal Fermi gas at zero temperature is given by the step function, $n(k)_{\rm IFG} = \left(\hbar n\right)^{-1}$ for $|k|<k_F$ ($k_F = \pi  n$ is the Fermi momentum) and zero elsewhere. The corresponding OBDM is given by its Fourier transform and is an oscillating function, $g_1(x)_{\rm IFG} =n \hbar \int_{-\infty}^{+\infty} n(k)_{\rm IFG} \cos(kx) dk/(2\pi) = n \sin(k_F x)/(k_F x)$, where the maxima of the Friedel oscillations occur at multiples of $k_F^{-1}$. 
Also, the fermionic OBDM changes its sign (for example, for a displacement which interchanges two nearest fermions,  $n^{-1}<x<2 n^{-1}$). Obviously, the OBDM for bosons is different, Fig.~\ref{fig:g110}, as $g_1(x)$ cannot be negative in the bosonic case, and a significant effort has been devoted to the calculation of the OBDM of the Tonks-Girardeau gas
\cite{Schultz1963,Lenard1964,VaidyaTracy79,VaidyaTracyErratum,Jimbo1980,Olshanii2003}.
Eventually, all coefficients of the zero-temperature short-range expansion, Eq.~\eqref{Eq:OBDM small x}, have been calculated analytically in Ref.~\cite{Astrakharchik2006II}.
Even at low temperature (upper panel of Fig.~\ref{fig:g110}), the coherence is rapidly lost, as $g_1(x)\propto 1/\sqrt{x}$.
The loss of coherence is exponentially fast at a finite temperature with the Fermi energy being the only relevant scale in the $\gamma\to\infty$ limit.
It is interesting to note, that the Maxwell-Boltzmann regime, Eq.~\eqref{Eq:g1 gaussian}, is reached only when the temperature is of the order of thousands of the quantum degeneracy temperature, see the lower panel of Fig.~\ref{fig:g110}. 

\section{Experimental considerations}
\label{Sec:experiments}

Cold atoms provide a powerful and clean platform for studying the physics of correlations in a widely tunable range of interaction and temperature. One-dimensional geometry can be realized in highly anisotropic trap configurations, where the confinement in the two radial $y$ and $z$ spatial directions is so strong that the motion of atoms is restricted to the $x$-axis. A reliable description of the system is then based on an effective 1D model, where both the temperature $T$ and the chemical potential $\mu$ are much smaller than the first excited energy level of the trapping potential \cite{Kruger2010}. This condition can be expressed as $k_B T, \mu \ll \hbar \omega_\perp$ where $\omega_\perp = \omega_y = \omega_z$ denotes the harmonic trap frequency in the strongly confined directions.

Highly anisotropic geometries are created in optical dipole traps \cite{Dettmer2001}, two-dimensional optical lattices generating a set of 1D tubes \cite{Greiner2001, Paredes2004, Tolra2004, Kinoshita2004, Kinoshita2005, Kinoshita2006, Vogler2013, Meinert2015, Fabbri2015}, on an atom chip \cite{Esteve2006, vanAmerongen2008, Kruger2010, Gring2012} and with a single optical tube trap \cite{Salces-Carcoba2018}. The latter two setups allow for the realization of a single 1D Bose gas which avoids the issue of ensemble averaging over many non-identical copies of the system taking place in lattice configurations. In addition, the exploration of even the classical gas regime of high temperature has been achieved in the optical tube trap \cite{Salces-Carcoba2018}, by keeping satisfied the condition $k_B T, \mu \ll \hbar \omega_\perp$ for the 1D geometry.

An important advantage of the one-dimensional geometry is that three-body losses are strongly suppressed \cite{Tolra2004}, making it possible to reach the limit of a diverging coupling constant (unitarity). 
In three dimensions, unitary Bose gas is instead prone to instability due to the formation of Efimov trimers, which are absent in one dimension. A spatial uniform density can be achieved in a flatbox potential \cite{Rauer2018} which can be advantageous as compared to the measurements in the axially trapped configuration where the density is dependent on the spatial coordinate and, for example, the simple power law of the OBDM in the Luttinger Liquid regime gets totally blurred by the trap. The interaction strength $\gamma \sim 1/\left(n a\right)$ can be tuned at will in experiments by adjusting the linear density $n$ via the strongly confining potential in the radial directions \cite{Olshanii1998, Kinoshita2004, Kinoshita2005} or by applying Fano-Feshbach resonances to tune the 3D scattering length, and correspondingly the 1D length $a$ in the Confinement Induced Resonance \cite{Chin2010, Meinert2015}.

The measurement of temperature (thermometry) of one-dimensional Bose gases can be achieved in various ways, which are often based on the comparison between experimental measurement statistics and theoretical predictions. For example, the temperature can be extracted from the pair correlation function of a single quasicondensate which is released and expands in the time-of-flight experiment along the 1D axis \cite{Imambekov2009, Manz2010}. As a result, many repetitions of the same measurement are needed in order to obtain an accurate comparison with the theory. Instead, thermometry from a single absorption image measurement during the time-of-flight expansion has been recently made possible with neural network \cite{Moller2021}. Compared to the standard method based on fitting the pair correlation function, the network can achieve the same precision needing only half the amount of experimental images.

In a 1D Bose gas, the pair correlation function $g_2\left(x\right)$ can be detected using spatially resolved in-situ single-atom counting as proposed in Ref.~\cite{Sykes2008}, standard absorption imaging \cite{Manz2010} and observing intensity correlations in the interference pattern of two identical spatially displaced quasicondensates \cite{Hellweg2003,Cacciapuoti2003} during time-of-flight expansion. The local value $g_2\left(0\right)$ has been measured by photoassociation \cite{Kinoshita2005}. The OBDM $g_1\left(x\right)$ has been detected with the matter-wave interferometry of two copies of expanding quasicondensates \cite{Hugbart2005}.

Beyond the implementation in ultracold atoms, the 1D Bose gas has been experimentally realized in superconducting circuits where the OBDM and the pair correlation functions have been measured \cite{Eichler2015}.

\section{Conclusions}
\label{Sec:conclusions}

In conclusion, we have carried out a detailed study of correlation functions in a one-dimensional Bose gas with contact interactions at finite temperatures.
The Path Integral Monte Carlo method is employed to perform exact calculations of the one- and two-body correlation functions, and the static structure factor. We find a good agreement with the predictions of a variety of perturbative and effective descriptions, as well as exact theories in regions of their applicability.

In particular, the short-range correlations have been extracted from several thermodynamic properties evaluated with Thermal Bethe Ansatz. The value of the pair correlation function $g_2\left(0\right)$ is related to the probability of two particles to overlap. It shows a competition of two opposite phenomena, that is the antibunching, i.e. tendency of two repulsive atoms to avoid each other, and a bunching effect, that is an enhanced probability of being close to each other due to thermal fluctuations. 
The short-range expansion of the One-Body Density Matrix contains both analytic and nonanalytic terms.
The analytic contributions ($x^n$) are related to $n$-th moments of the momentum distribution and have non-vanishing coefficients for even powers, the leading one ($x^2$) can be expressed in terms of the average kinetic energy.
The nonanalytic contributions come as an expansion of powers of $|x|$, being the leading one proportional to $|x|^3$ whose coefficient depends on the Tan's contact and thus captures the amplitude of the $1/k^4$-decay of the momentum distribution. By comparison with PIMC results, we show that the short-range expansion of the OBDM is valid for any value of the interaction strength and temperature. 

At zero temperature, the coherence is lost in a slow long-range power-law decay for the OBDM allowing for the application of condensate-based theories (Gross-Pitaevskii, Bogoliubov, etc.)
in the weakly-repulsive regime. At temperature higher than the hole-anomaly threshold, the Decoherent Classical perturbative theory well reproduces PIMC results for the pair correlation function. It captures both the bunching effect, described by a Gaussian term whose width is provided by the de Broglie thermal wavelength, and the antibunching of strength proportional to the interaction parameter $\gamma$, Eq.~\eqref{Eq:g2 DC}.
By further increasing the temperature, correlations are no longer affected by Bose or Fermi statistics but rather are described by the Maxwell-Boltzmann Gaussian decay which is independent on $\gamma$.

The low-momentum behavior of the static structure factor is well captured by a single-mode quasiparticle approximation below the hole-anomaly temperature.
A linear-phonon single-mode dispersion lies in the base of the Luttinger Liquid theory whose applicability is confirmed for any interaction strength.
In the weakly-repulsive regime, the quasiparticle excitation is provided by the Bogoliubov spectrum which is linear at small momentum but deviates at larger momenta. 
The zero-momentum static structure factor $S(0)$ is related to the isothermal compressibility and its value can be extracted from the Thermal Bethe Ansatz. It quantifies the interplay between the quantum and thermal fluctuations and shows that the thermal effects are more important for weak interactions. 

Our exact PIMC and thermal Bethe-Ansatz results do not agree with some known analytical limits of the pair correlation function $g_2\left(x\right)$ \cite{Sykes2008, Deuar2009} and of its local value $g_2\left(0\right)$ \cite{Kheruntsyan2003}, respectively, the latter of which is a thermodynamic property. Improved analytical descriptions have been recently developed and they exhibit an excellent matching with the exact TBA method for several thermodynamic quantities \cite{DeRosi2017, DeRosi2019, DeRosi2021II}.

We expect that our PIMC calculations of the correlation functions will serve as a universal benchmark for future theories and experimental measurements. Our results may stimulate further theoretical and experimental investigations aimed at the complete understanding at the microscopic level of 1D Bose gases through quantum many-body correlations which allow for the identification of different quantum regimes. The PIMC results for the momentum distribution will be presented in an upcoming publication \cite{DeRosi2023}. 

Looking forward, interesting perspectives include the calculation of $n$-body correlations \cite{Bastianello2018} and the study in optical lattices \cite{Rigol2005}.
Particularly appealing is the extension to the harmonically axially-trapped configuration \cite{Petrov2000, Kheruntsyan2005, Davis2012, Vignolo2013, Xu2015, Yao2018, Minguzzi2022} which is suitable for the investigation of breathing modes \cite{Moritz2003, Hu2014, Fang2014} whose collective frequencies change by crossing different quantum regimes \cite{DeRosi2015}. Recently, a temperature-induced hydrodynamic-to-collisionless transition has been predicted with the emergence of the excitation of two different frequencies in the time-evolution of breathing modes in both the GP and TG limits \cite{DeRosi2016}.
Such a transition may shed light on the absence of thermalization in 1D Bose gases even after thousands of interatomic collisions \cite{Kinoshita2006}.

Our results can be extended to systems with positive 1D $s-$wave scattering length $a > 0$: (i) gases with short-range interactions like the strongly correlated metastable state (super Tonks-Girardeau gas) \cite{Astrakharchik2005, Haller2009, Kormos2011II} and the hard-core model \cite{Tonks1936, Mattis1993, Wadati2005, Mazzanti2008, Motta2016}, (ii) finite-range interacting systems such as dipolar \cite{Arkhipov2005,Citro2007} and Rydberg atoms \cite{Osychenko2011}, $^{4}$He liquid \cite{Bertaina2016}, $^{3}$He gas \cite{Astrakharchik2014}, and a single chain of $^{4}$He atoms confined in a nanopore even with disorder \cite{Vranjes2018}.
Other interesting extensions of our work include spin models \cite{Deuretzbacher2016, Labuhn2016} and multicomponent
systems \cite{Gharashi2013, Yang2015, Decamp2016, Patu2017}.
Our predictions are important for the exploration of the properties of impurities immersed in helium \cite{Bardeen1967} and in a Bose gas \cite{Reichert2019, Pascual2021} as a function of the interaction strength of the bath and temperature.
Possible extensions of our work also include quantum liquids in binary bosonic mixtures at finite temperature and 1D geometry \cite{DeRosi2021} which can be realized in current experiments \cite{Cheiney2018}.
Finally, our findings can be also applied to 1D liquids of $^4$He which have been recently observed \cite{DelMaestro2022} in a wide range of interaction and temperature regimes. 

\begin{acknowledgments}

G. D. R.'s received funding from the European Union's Horizon 2020 research and innovation program under the Marie Sk\l odowska-Curie grant agreement {\it UltraLiquid} No. 797684 and with the grant IJC2020-043542-I funded by MCIN/AEI/10.13039/501100011033 and by "European Union NextGenerationEU/PRTR". G. D. R., G. E. A. and J. B. were partially supported by grant PID2020-113565GB-C21 funded by MCIN/AEI/10.13039/501100011033 and the Spanish MINECO (FIS2017-84114-C2-1-P).
\end{acknowledgments}

\appendix

\section{Coefficient $c_2$ of the OBDM at small distances}
\label{Sec:c2}
Simple considerations on scale invariance \cite{FetterBook, Barth2011} lead to the exact thermodynamic relation holding for any value of interaction strength and temperature \cite{DeRosi2019}:
\begin{equation}
\label{Eq:scale inv}
- \frac{\mathcal{C}}{N} \frac{\hbar^2 a}{4 m} = 2 \frac{E}{N} - \frac{P}{n} .
\end{equation}
Eq.~\eqref{Eq:scale inv} relates the Tan's contact, Eq.~\eqref{Eq:contact}, the pressure, Eq.~\eqref{Eq:Pressure} and the total internal energy $E = \langle H_{\rm kin} \rangle + \langle H_{\rm int} \rangle$ where the average interaction contribution can be calculated from the Hellmann-Feynman theorem \cite{Feynman1939, DeRosi2016}:
\begin{equation}
\label{Eq:Hint}
\langle H_{\rm int} \rangle = g \frac{\partial A}{\partial g} = - \mathcal{C} \frac{\hbar^2 a}{4 m} \ ,
\end{equation}
and where we have used Eq.~\eqref{Eq:contact}.
By combining Eqs.~\eqref{Eq:scale inv}-\eqref{Eq:Hint}, one calculates the average kinetic energy:
\begin{equation}
\label{Eq:Hkin}
\langle H_{\rm kin} \rangle = \frac{N}{2} \left(\frac{P}{n} - \frac{\langle H_{\rm int} \rangle }{N} \right) \ .
\end{equation}

By using Eq. \eqref{Eq:Hint} in Eq.~\eqref{Eq:c2} of the main text, one finally finds \cite{Pitaevskii2016}:
\begin{equation}
c_2 = - \frac{m}{\hbar^2 n^2}  \frac{\langle H_{\rm kin} \rangle}{N} = -\frac{1}{2} \frac{\langle k^2 \rangle}{n^2}\ ,
\end{equation}
where the average kinetic energy per particle can be expressed as $\langle H_{\rm kin} \rangle/N = \hbar^2 \langle k^2\rangle/\left(2 m\right)$ with $\langle k^2 \rangle = N^{-1} \hbar \int_{-\infty}^{+\infty} dk  n(k)k^2$ calculated from the momentum distribution $n\left(k\right)$ which is a function of the momentum $k$ \cite{Pitaevskii2016}.

\bibliography{Bibliography}

%%%%%%%%%%%%%%%%%%%%%%%%%%%%%%%%%%%%%%%%%%%%%

 \renewcommand{\theequation}{S\arabic{equation}}
 \setcounter{equation}{0}
 \renewcommand{\thefigure}{S\arabic{figure}}
 \setcounter{figure}{0}
 \renewcommand{\thesection}{S\arabic{section}}
 \setcounter{section}{0}
 \onecolumngrid

\end{document}